\documentclass{spie}
\usepackage[latin9]{inputenc}
\usepackage{color}
\usepackage{float}
\usepackage{textcomp}
\usepackage{amsmath}
\usepackage{amssymb}
\usepackage{graphicx}
\usepackage{esint}
\usepackage[unicode=true,
 bookmarks=false,
 breaklinks=false,pdfborder={0 0 1},backref=section,colorlinks=true]
 {hyperref}
\hypersetup{
 allcolors=blue}

\makeatletter

\DeclareFontEncoding{LGR}{}{}

\ProvideTextCommand{\~}{LGR}[1]{\char126#1}

\def\be{\begin{equation}}
\def\ee{\end{equation}}
\def\ber{\begin{eqnarray}}
\def\eer{\end{eqnarray}}

\newcommand{\commentout}[1]{}

\usepackage{amsfonts}

\title{Theory of charge-spin conversion at oxide interfaces: \\The inverse spin-galvanic effect}

\author[a]{G\"otz Seibold}
\author[b]{Sergio Caprara}
\author[c]{Roberto Raimondi}
\affil[a]{Institut f\"ur Theoretische Physik, BTU Cottbus-Senftenberg, PBox 101344, 03013 Cottbus, Germany}
\affil[b]{Dipartimento di Fisica Universit\`a di Roma Sapienza, piazzale Aldo Moro 5, I-00185 Roma, Italy}
\affil[c]{Dipartimento di Matematica e Fisica, Universit\`a Roma Tre,
Via della Vasca Navale 84, 00146 Rome, Italy}

\authorinfo{Further author information: (Send correspondence to Roberto Raimondi.)\\Roberto Raimondi: E-mail: roberto.raimondi@uniroma3.it}

\makeatother

\begin{document}
\maketitle 
\begin{abstract}
We evaluate the non-equilibrium spin polarization induced by an applied
electric field for a tight-binding model of electron states at oxides
interfaces in LAO/STO heterostructures. By a combination of analytic
and numerical approaches we investigate how the spin texture of the
electron eigenstates due to the interplay of spin-orbit coupling and
inversion asymmetry determines the sign of the induced spin polarization
as a function of the chemical potential or band filling, both in the
absence and presence of local disorder. With the latter, we find that
the induced spin polarization evolves from a non monotonous behavior
at zero temperature to a monotonous one at higher temperature. Our
results may provide a sound framework for the interpretation of recent
experiments.
\end{abstract}


\section{INTRODUCTION}

\label{sec:intro}

It is well known that the breaking of the inversion symmetry leads
to the so-called Rashba spin-orbit coupling (SOC)\cite{Rashba1959,Bychkov1984a,Bychkov1984},
where polar and axial vectors transform similarly\cite{Ganichev2016}.
Basically this allows for two major possibilities of charge to spin conversion: The spin Hall (SH) \cite{djakperel71} and the
inverse spin galvanic (ISG) effect \cite{Ivchenko1978,Vorobev1979}, as well as for their Onsager reciprocal effects.
While the SH effect converts an electrical current into a spin imbalance
at the sample edges via an induced perpendicular spin current, the
ISG effect creates a {\it bulk} non-equilibrium spin polarization
by a flowing electrical current\cite{Edelstein1990,Ivchenko1978,Ivchenko1989,Aronov1989,Levitov1985}. The inverse SG effect corresponds then to the
production of electrical current via the pumping of spin
polarization\cite{Ganichev2001,Ganichev2002}.
Both the SG \cite{Ganichev2001,Ganichev2002} and the ISG \cite{Kato2004,Kato2004b,Yang2006,Chang2007,Norman2014,Sih2005,Luengo2017}
effects have been observed in semiconductors. In the first case an
electrical current is measured after pumping spin polarized light
(SG) whereas in the second case Faraday and Kerr spectroscopies measure
the spin polarization induced by the applied current. The SG effect
has also been very effectively measured by spin pumping from an adjacent
ferromagnet into a metallic interface\cite{Sanchez2013}, into a topological
insulator surface\cite{Shiomi2014,Mellnik2014} and more recently
into the two-dimensional electron gas (2DEG) in oxide LAO/STO heterostructures\cite{Lesne2016,Chauleau2016,Song2017,Wang2017}.
These latter materials have emerged\cite{Soumyanarayanan2016,SC2016,Ando2017,Varignon2018,WeiHan2018}
as very promising materials for the SG and ISG effect, due to the
large values of the Rashba SOC parameter $\alpha$ as experimentally
observed\cite{Caviglia2010,Hurand2015,Gopinadhan2015,Liang2015} and
also theoretically calculated\cite{Shanavas2014,zhonh13,Bucheli2014},
even though it is likely that, due to their complex band structure,
the available theory\cite{Shen2014} of the SG/ISG effect developed
for the 2DEG in semiconductors may not be able to capture a number
of specific features. A first step in this direction has been made
recently by a combination of analytical diagrammatic and numerical
approaches\cite{Seibold2017,Sahin2018}.

The layout of the paper is the following. In the next section we introduce
a model for the electron states relevant for describing transport
at oxide LAO/STO interfaces. In section 3 we provide the necessary
formalism of linear response theory for the SG and ISG effects. In
section 4 we introduce an approximate effective model for electron
states close to band minima. In section 5 we evaluate analytically
the SG response for the effective model, whereas in section 6 we introduce
disorder and the necessary formalism to handle it. Finally in section
7 we present a fully numerical approach which includes both cases without
and with disorder. We conclude in section 8. A number of technical
details are provided in the appendices.

\section{THE MODEL}

\label{sec:model}

The electronic structure of the 2DEG at LAO/STO interfaces, perpendicular
to the $(001)$ crystal direction, is usually described \cite{zhonh13,Khalsa2013,Kim2013}
within a tight-binding Hamiltonian $H_{0}$ for  the Ti t$_{2g}$ orbitals, $d_{xy},\:d_{xz,}\:d_{yz}$, supplemented by local atomic spin-orbit
interactions with Hamiltonian $H_{aso}$ and an interorbital hopping
with Hamiltonian $H_{I}$ which is induced by the interface asymmetry.

The hopping between the $d$ orbitals of two neighbouring cubic cells
is mediated via intermediate jumps to $p$ orbitals. For instance,
the hopping between two $d_{xy}$ orbitals along the x axis occurs
via two successive hopping $d_{xy}\rightarrow p_{y}$ and $p_{y}\rightarrow d_{xy}$.
In the first hop, the overlap, which is of order $\sim t_{pd}$, yields
a positive sign, whereas the sign is negative $\sim-t_{pd}$ in the
second one. Hence the effective $d_{xy}-d_{xy}$ hopping goes like
$-t_{pd}^{2}/\Delta E$, with $\Delta E$ being the energy difference
between $d$ and $p$ orbitals. As a result, in the basis ${|xy\rangle,|xz\rangle,|yz\rangle}$
the hopping between similar orbitals reads as 
\begin{equation}
H_{0}=\left(\begin{array}{ccc}
\varepsilon_{k}^{xy} & 0 & 0\\
0 & \varepsilon_{k}^{xz} & 0\\
0 & 0 & \varepsilon_{k}^{yz}
\end{array}\right)\label{eq:hamiltonian0}
\end{equation}
with, setting to unity the lattice spacing, 
\begin{eqnarray*}
\varepsilon_{k}^{xy} & = & -2t_{1}\lbrack\cos(k_{x})+\cos(k_{y})-2\rbrack-4t_{3}\lbrack\cos(k_{x})\cos(k_{y})-1\rbrack\\
\varepsilon_{k}^{xz} & = & -2(t_{1}+t_{3})\lbrack\cos(k_{x})-1\rbrack-2t_{2}\lbrack\cos(k_{y})-1\rbrack+\Delta\\
\varepsilon_{k}^{yz} & = & -2(t_{1}+t_{3})\lbrack\cos(k_{y})-1\rbrack-2t_{2}\lbrack\cos(k_{x})-1\rbrack+\Delta
\end{eqnarray*}
where the energy difference $\Delta$ between the $|xy\rangle$ and
$|xz\rangle,|yz\rangle$ states is due to the confinement of the 2DEG
in the $xy$-plane\cite{Scopigno2016}.

The atomic SOC is given by 
\begin{equation}
H_{aso}=\Delta_{aso}\left(\begin{array}{ccc}
0 & -i\tau^{x} & i\tau^{y}\\
i\tau^{x} & 0 & -i\tau^{z}\\
-i\tau^{y} & i\tau^{z} & 0
\end{array}\right)\label{eq:hatso}
\end{equation}
with $\tau^{i}$ denoting the Pauli matrices.

\begin{figure}[H]
\begin{center}
\includegraphics[width=12cm]{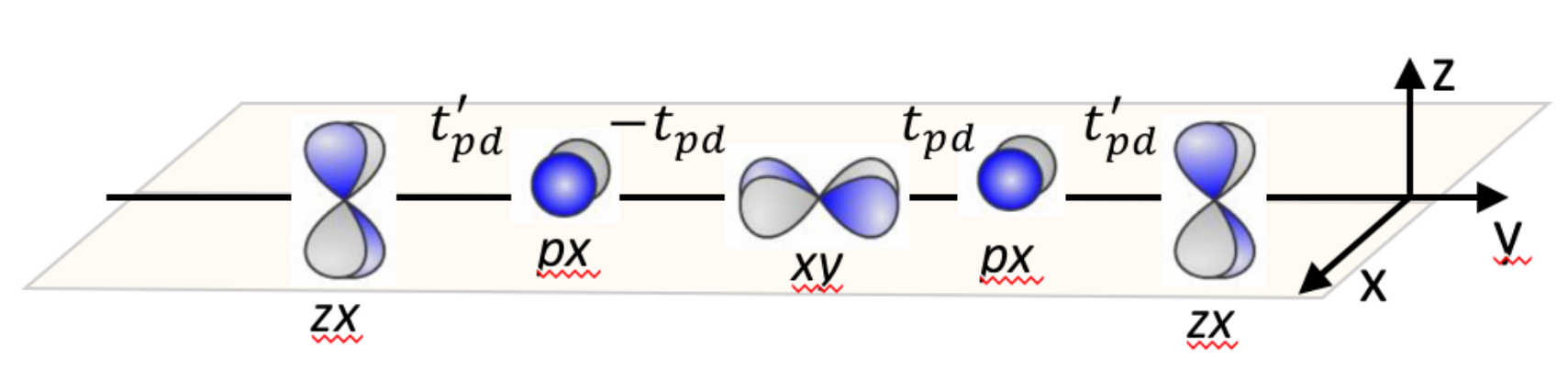}
\end{center}
\caption{At the interface an orbital polarization
and (or) orbital displacement results in hopping processes $\sim t'_{pd}$
as between $p_x$- and $zx$-orbitals along the y-direction.
The asymmetry is visualized by a small shift of $p_x$-orbitals
along the z-direction. \label{fig:asymhop}}
\end{figure}

\begin{figure}
\includegraphics[width=17cm]{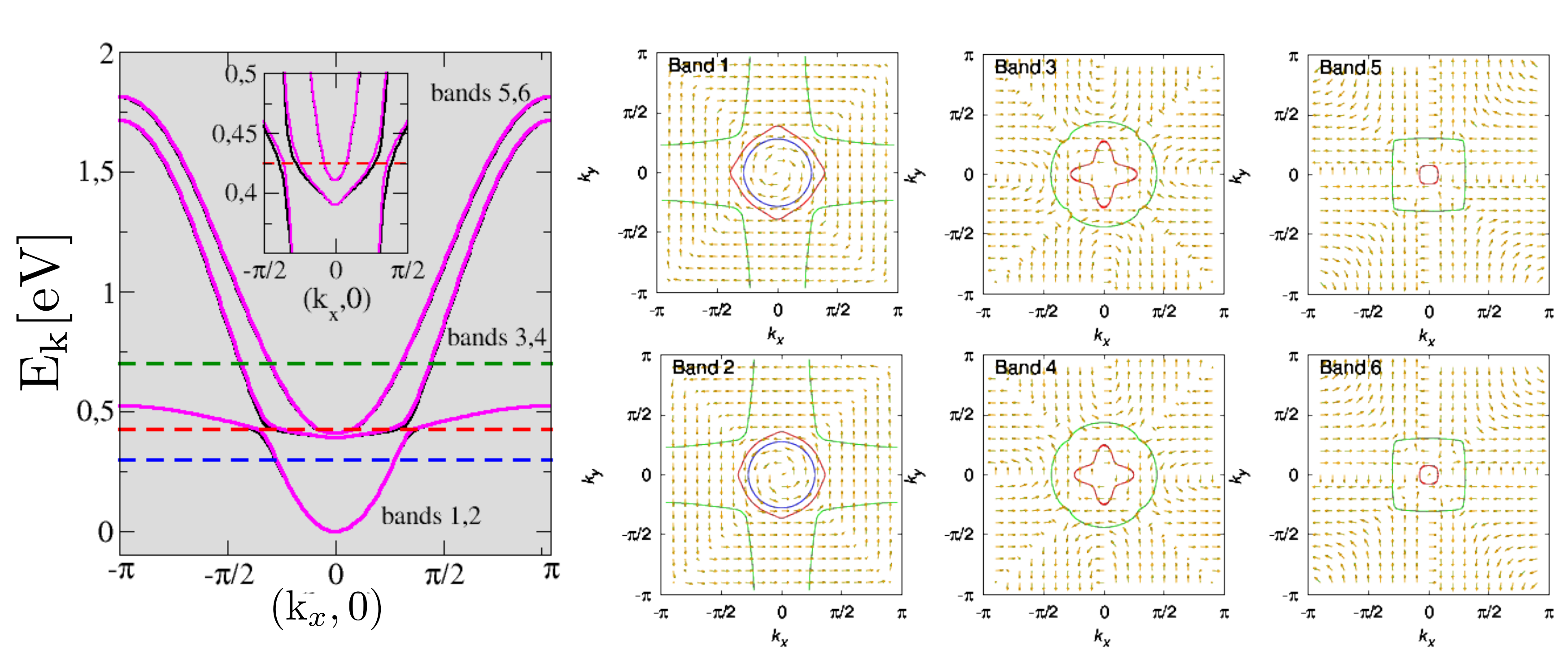}
\caption{Left panel: Structure of the $t_{2g}$ interface bands. The inset
enlarges the region around the 'Lifshitz' point where the spin-orbit
splitting is large. The horizontal dashed lines in the main panel
refer to three values of chemical potential: $\mu=0.3$ eV (blue line),
$\mu=0.425$ eV (red line) and $\mu=0.7$ eV (green line). The right
panel displays the spin texture for the three pairs of bands together
with their Fermi surfaces. For the lower pair of bands (1,2) the spins
point in the opposite direction.\label{fig:chiralities}}
\end{figure}

Hopping between different $d$ orbitals may occur if inversion symmetry
is broken, see Fig. \ref{fig:asymhop}. Consider, for instance, the
two hops along the y direction,
$d_{xy}\rightarrow p_{x}$ and $p_{x}\rightarrow d_{zx}.$ While the
first hop $\sim t_{pd}$ is as the first hop of the effective hopping
between two $d_{xy}$ orbitals discussed above, the second hop $\sim t_{pd}^{'}$
will be forbidden in the presence of inversion symmetry. To see this
consider that

\[
t_{pd}=\langle p_{x},\mathbf{R}+\frac{a}{2}\mathbf{y}\mathrm{|}H|d_{xy},\mathbf{R}\rangle,
\]
where $\mathbf{R}$ is the lattice site of the $d_{xy}$ orbital.
In the same way

\[
t_{pd}^{'}=\langle d_{zx},\mathbf{R}+a\mathbf{y}|H|p_{x},\mathbf{R}+\frac{a}{2}\mathbf{y}\rangle.
\]
In both cases, $H$ is the full Hamiltonian. If $H$ is invariant
with respect to the inversion $z\rightarrow-z$, then necessarily
$t_{pd}^{'}=0$, because $p_{x}$ is even, while $d_{zx}$ is odd.
Clearly if $H$ has terms which are not invariant for $z\rightarrow-z$,
then $t_{pd}^{'}\neq0$. As a result the interface asymmetry hopping
reads\cite{Khalsa2013} 
\begin{equation}
H_{I}=\gamma\left(\begin{array}{ccc}
0 & -2i\sin(k_{y}) & -2i\sin(k_{x})\\
2i\sin(k_{y}) & 0 & 0\\
2i\sin(k_{x}) & 0 & 0
\end{array}\right)\,.\label{eq:ashop}
\end{equation}
In the following we use the parameters, $t_{1}=0.277$\,eV, $t_{2}=0.031$\,eV,
$t_{3}=0.076$\,eV, $\Delta=0.4$\,eV, $\Delta_{aso}=0.010$\,eV,
$\gamma=0.02$\,eV, which have been derived in Ref. \cite{zhonh13}
from projecting DFT on the $t_{2g}$ Wannier states. Note that for
the splitting $\Delta$ we take a value intermediate between the theoretical
($\Delta=0.19$\,eV) and the experimental one ($\Delta=0.6$\,eV).
The left panel of Fig. \ref{fig:chiralities} shows the band dispersions
along the x axis for these values of the parameters. The bands come
naturally in three pairs, which are split by the combined effect of
the spin-orbit coupling and the inversion symmetry breaking. For our analysis
we have selected three different values of the chemical potential
for corresponding filling regimes. For $\mu=0.3$ eV, only the lowest
pair of bands (1,2) is occupied. The chemical potential $\mu=0.425$
eV is close to the Lifshitz point, where the spin-orbit splitting
is large and the pairs of bands (3,4) and (5,6) start to be filled.
Finally, the chemical potential $\mu=0.7$ eV is in the regime, where
all pairs of bands (1,2), (3,4) and (5,6) are occupied.

We now analyze the chirality for each eigenstate band $p=1,\dots,6$
by computing the spin at each momentum point of the Fermi surface
(FS) according to

\[
S^{\alpha}(p,k_{F})=\sum_{n,\sigma,\sigma'}\Phi_{n,\sigma}^{*}(p,k_{F})\tau_{\sigma,\sigma'}^{\alpha}\Phi_{n,\sigma'}(p,k_{F})
\]
where $\Phi_{n,\sigma'}(p,k_{F})$ are the eigenfunctions of the system
at momentum $k_{F}$. The indices $n=xy,xz,yz$ and $\sigma$ label
the orbital and its spin. Then the chirality of the $p$-th band can
be obtained from

\begin{equation}
\nonumber
\alpha(p)=\arcsin \left(\frac{{\bf k}_{F}\times{\bf S}(p,k_{F})\cdot {\bf e}_{z}}{|{\bf k}_{F}||{\bf S}(p,k_{F}|}\right) \,.
\end{equation}
Fig.\,\ref{fig:chiralities} shows the chiralities for each pair
of bands at selected chemical potentials and the corresponding FSs.
For the lowest pair of bands (1,2) the momentum dependent
spin pattern displays a vortex-type structure with the core centered at
$\Gamma=(0,0)$. Thus, even when the FS changes from electron- to
hole-like between $\mu=0.5$ and $\mu=0.6$, the corresponding
chiralities are always confined to $\alpha(p)\approx\pm\pi/2$ without
any sign change in $\alpha$. For the middle pair of bands (3,4)  the spin structure is composed of two vortex patterns (with the same vorticity) centered at $(\pi,0)$ and $(0,\pi)$. As a consequence, the spin texture
vanishes along the diagonals and a Rashba-type description along this direction fails.
In section \ref{sec:effectivemodels} we will come back to this point.
However, for small  $\mu$ and all other momenta the chirality
also starts at $\alpha\approx\pm\pi/2$ but
then on average becomes smaller with increasing chemical potential
and eventually changes sign for $\mu\approx 1.5$.  An analogous
situation occurs for the uppermost pair of bands where the 'spin-vortex core'
is centered at $(\pi,\pi)$. In this case  the chiralities
also change sign upon increasing the chemical
potential while at small $\mu$ one again recovers
$\alpha\approx\pm\pi/2$.

\section{LINEAR RESPONSE THEORY}

\label{sec:LRT} In this paper we aim at evaluating the spin polarization
induced by an externally applied electric field. To be definite we
take the electric field along the x axis and the spin polarization
along the y axis. To linear order in the applied field we write the
spin polarization as

\begin{equation}
s^{y}(\omega)=\sigma^{ISG}(\omega)\,E_{x}(\omega),\label{eq:LR_ISG}
\end{equation}
where $\sigma^{ISG}$, the ``conductivity'' for the ISG effect,
can be obtained by the zero-momentum limit of the Fourier transform
$R_{yx}(\omega)$ of the response function (henceforth the symbols
in capital letters indicate the operators for spin density and charge
current) defined as

\begin{equation}
R_{yx}(t,\mathbf{r})=-\imath\theta(t)\left\langle \left[\mathrm{S}^{y}(t,\mathbf{r}),\mathrm{J}_{x}\right]\right\rangle ,\label{eq:LinRes}
\end{equation}
where the brackets stand for the quantum-statistical average and $\theta\left(t\right)$
is the Heaviside step function. The frequency-dependent ISG conductivity
reads

\begin{equation}
\sigma^{ISG}(\omega)=\lim_{\eta\rightarrow0^{+}}\Re\left[\frac{R_{yx}(\omega)}{\imath(\omega+\imath\eta)}\right]=-\pi\delta(\omega)R_{yx}^{'}(0)+\mathcal{P}\frac{R_{yx}^{''}(\omega)}{\omega}\equiv D^{ISG}\delta\left(\omega\right)+\mathcal{P}\frac{R_{yx}^{''}(\omega)}{\omega},\label{eq:FD_cond}
\end{equation}
where the first term will be referred to as the Drude singular term
and the second as the regular term, in analogy with the terminology
used in the case of the optical conductivity. Because under time reversal
both the charge current and the spin polarization are odd, according
to the Onsager relation, the SG and ISG conductivities are equal\cite{Shen2014}.
For this reason we will use the term SG conductivity (SGC) for both
direct and inverse effects. The calligraphic symbol $\mathcal{P}$
stands for the principal part. The real $R_{yx}^{'}(\omega)$ and
the imaginary $R_{yx}^{''}(\omega)$ parts of the response function
are related by the Kramers-Kronig relation (KKR)

\begin{equation}
R_{yx}^{'}(\omega)=\frac{1}{\pi}\fint_{-\infty}^{\infty}d\,\omega^{'}\frac{R_{yx}^{''}(\omega^{'})}{\omega^{'}-\omega}.\label{eq:KK}
\end{equation}
By integration over the frequency, thanks to the KKR, the SGC satisfies
the following sum rule

\begin{equation}
\int_{-\infty}^{\infty}d\,\omega\,\sigma^{ISG}(\omega)=0\label{eq:sum_rule}
\end{equation}
due to the fact that for the SGC there is no 'diamagnetic' contribution as opposed to the optical conductivity.

In the following we are going to apply the above formulae to the model
introduced in section \ref{sec:model}. To this end, it is instructive
to consider first the case of the Rashba SOC for a 2DEG with quadratic
dispersion relation in the effective mass approximation. The insight
gained in this simpler case will guide us also in the analysis of
the model with a complex band structure. We consider then the Rashba-Bychkov
Hamiltonian\cite{Bychkov1984}

\begin{equation}
H=\frac{p^{2}}{2m}+\alpha(\tau^{x}p_{y}-\tau^{y}p_{x}),\label{eq:Rashba2DEG}
\end{equation}
where $m$ is the effective mass and $\alpha$ the SOC. The 2DEG is
confined to the xy plane and $p_{x}$ and $p_{y}$ are the momentum
operators along the two coordinate axes. Clearly there are two eigenvalues
$E_{\pm}(p)=p^{2}/2m\pm\alpha p$ with the corresponding eigenstates
of (\ref{eq:Rashba2DEG}) being plane waves whose spin quantization
axis is fixed by the momentum direction

\begin{equation}
|\mathbf{p},s\rangle=\frac{1}{\sqrt{2}}\left(\begin{array}{c}
s\,\imath\mathrm{e}^{-\imath\theta}\\
1
\end{array}\right),\:\;s=\pm1\label{eq:RashbaEIGEN}
\end{equation}
where $\tan(\theta)=p_{y}/p_{x}.$ The ISG response function at finite
frequency and momentum reads

\begin{equation}
R_{yx}(\omega,\mathbf{q})=\sum_{\mathbf{p},s_{1},s_{2}}\left\langle \mathbf{p},s_{1}|\mathrm{S}^{y}|\mathbf{p},s_{2}\right\rangle \left\langle \mathbf{p},s_{2}|\mathrm{J}_{x}|\mathbf{p},s_{1}\right\rangle \frac{\,f(E_{s_{1}}(\mathbf{p})-\mu)-f(E_{s_{2}}(\mathbf{p}+\mathbf{q})-\mu)}{\omega+\imath\eta+E_{s_{1}}(\mathbf{p})-E_{s_{2}}(\mathbf{p}+\mathbf{q})},\label{eq:Response2DEG}
\end{equation}
where $f\left(E\right)$ is the Fermi distribution function at temperature
$T$. Depending on the values of the spin indices, one has intraband
($s_{1}=s_{2}=\pm1$) and interband ($s_{1}=-s_{2}=\pm1$) contributions.
In the dynamic limit, when the momentum goes to zero at finite frequency,
the intraband contribution vanishes. For the model of Eq.\,(\ref{eq:Rashba2DEG})
the interband matrix elements for spin density $\mathrm{S}^{y}=\tau^{y}/2$
and charge current $\mathrm{J}_{x}=(-e)(p_{x}/m-\alpha\tau^{y})$
read

\begin{eqnarray*}
\left\langle \mathbf{p},s|\mathrm{S}^{y}|\mathbf{p},-s\right\rangle  & = & \frac{1}{2}(-\imath s)\sin(\theta),\\
\left\langle \mathbf{p},-s|\mathrm{J}_{x}|\mathbf{p},s\right\rangle  & = & (-e)\alpha(\imath s)\sin(\theta),
\end{eqnarray*}
and the zero-momentum response function becomes

\begin{equation}
R_{yx}(\omega)=\frac{1}{2}(-e)\alpha\sum_{\mathbf{p}\mathrm{s}}\sin^{2}(\theta)\frac{\,f(E_{s}(\mathbf{p})-\mu)-f(E_{-s}(\mathbf{p})-\mu)}{\omega+\imath\eta+E_{s}(\mathbf{p})-E_{-s}(\mathbf{p})}.\label{eq:ResponseZeroMomentum}
\end{equation}
At zero temperature, there are two FSs corresponding to the two spin
helicity bands with Fermi momenta $p_{\pm}=\sqrt{2m\mu+(m\alpha)^{2}}\mp m\alpha$.
The evaluation of the imaginary part of the zero-momentum response
function leads to ($\eta\rightarrow0^{+}$)

\begin{equation}
R_{yx}^{''}\left(\omega\right)=\frac{e}{32\alpha}\omega\left[\theta\left(\left|\omega\right|-2\alpha p_{-}\right)-\theta\left(\left|\omega\right|-2\alpha p_{-}\right)\right],\label{eq:imaginaryResponse}
\end{equation}
showing an antisymmetric behavior with respect to the frequency $\omega$.
The spectral weight, at positive frequency, is confined in the range
$2\alpha p_{+}<\omega<2\alpha p_{-}.$ The two frequencies delimiting
the interval are nothing but the spin-orbit splitting at the two Fermi
surfaces. We note, and this will turn out useful when discussing the
numerical calculations, that at finite $\eta$, the imaginary part
remains finite and acquires a linear-in-frequency behavior around
the origin, whose slope vanishes as $\eta$. The Drude weight, according
to Eq.\,(\ref{eq:FD_cond}) can be easily obtained by the KKR relation
(\ref{eq:KK}) to read

\begin{equation}
D^{ISG}=-\frac{\pi}{2}eN_{0}\alpha,\label{eq:Drude2DEG}
\end{equation}
where $N_{0}=m/(2\pi)$ is the single-particle density of states of
the 2DEG. For the sake of simplicity we have chosen units such $\hbar =1$. There are two features worth noticing. The first is that
the Drude weight is controlled by the sign of the SOC. The second
is that the Drude weight arises from the interband transitions between
the spin-orbit split bands. This must be compared with the case of
optical conductivity for the electron gas, where the Drude weight
arises from the diamagnetic contribution to the current. In the present
case, due to the sum rule (\ref{eq:sum_rule}), the Drude low-frequency
peak yields information about the spectral weight of interband transitions
at finite frequency. To the best of our knowledge this feature has
not been noticed before.

In the following of the paper we will consider the effect of disorder,
but it is instructive to make here an heuristic discussion. In the
presence of spin-independent disorder, due to the form (\ref{eq:RashbaEIGEN})
of the eigenstates, the electron spin acquires a finite relaxation
rate $\tau_{s}^{-1}$. This mechanism, which is known as the Dyakonov-Perel
relaxation, arises because, at each scattering event, the change in
momentum also affects the spin eigenstate. As a result, in the diffusive
approximation, $\omega\tau\ll1$, the spin density obeys a Bloch equation\cite{Raimondi2006}

\begin{equation}
\frac{{\rm d}s^{y}}{{\rm d}t}=-\frac{1}{\tau_{s}}(s^{y}-s_{0}),\label{EOM_spin}
\end{equation}
where $s_{0}=-e\alpha N_{0}\tau E$ represents the steady-state nonequilibrium
spin polarization\cite{Edelstein1990} induced by an applied electric
field $E$ along the x axis and $\tau$ is the momentum relaxation
scattering time (not to be confused with the Pauli matrices $\tau^{i}$).
According to Ref.\cite{Raimondi2006} the Dyakonov-Perel relaxation
rate reads

\begin{equation}
\frac{1}{\tau_{s}}=\frac{1}{2\tau}\frac{4\alpha^{2}p_{F}^{2}\tau^{2}}{1+4\alpha^{2}p_{F}^{2}\tau^{2}}.\label{DPSRT}
\end{equation}
By Fourier transforming (\ref{EOM_spin}) to frequency $\omega$,
one obtains the SGC in the form

\begin{equation}
\sigma^{ISG}(\omega)=-e\alpha N_{0}\frac{\tau}{\tau_{s}}\frac{\tau_{s}^{-1}}{\omega^{2}+\tau_{s}^{-2}},\label{eq:FDSISG}
\end{equation}
which has a Lorentzian lineshape and evolves to a singular contribution
in the weak scattering limit $\tau\rightarrow\infty$. More precisely
by integrating over frequency one obtains

\begin{equation}
\int_{-\infty}^{\infty}d\omega\sigma^{ISG}(\omega)=-e\alpha N_{0}\pi\frac{\tau}{\tau_{s}}=-\frac{\pi}{2}e\alpha N_{0},\label{eq:LorentzianDrude}
\end{equation}
which reproduces the Drude weight of Eq.\,(\ref{eq:Drude2DEG}).
Notice that in the last step we made use of the fact that the spin
relaxation time becomes twice the momentum relaxation time in the
weak scattering limit according to Eq.\,(\ref{DPSRT}). Eq.\,(\ref{eq:LorentzianDrude})
seems to violate the sum rule (\ref{eq:sum_rule}), but this is not
the case. The form (\ref{eq:FDSISG}) for the SGC has been derived
in the diffusive approximation, which is valid for frequencies $\omega\ll\tau^{-1}$
well below the region of the interband spectral weight. Hence, the
form (\ref{eq:FDSISG}) captures only the low frequency spectral weight,
which evolves in the singular Drude weight in the limit of vanishing
disorder. The effect of disorder is then to eliminate the Drude singular
contribution and to yield a finite SGC at zero frequency, which is
the result of a finite slope of the imaginary part of the response
function. The microscopic approach in the presence of disorder is
discussed in section \ref{sec:disorderedlimit} and details about
the frequency dependence are developed in the appendix \ref{sec:FFBS}.

\section{EFFECTIVE MODELS}

\label{sec:effectivemodels}

\begin{figure}[htb]
\begin{center}
\includegraphics[clip,width=12cm]{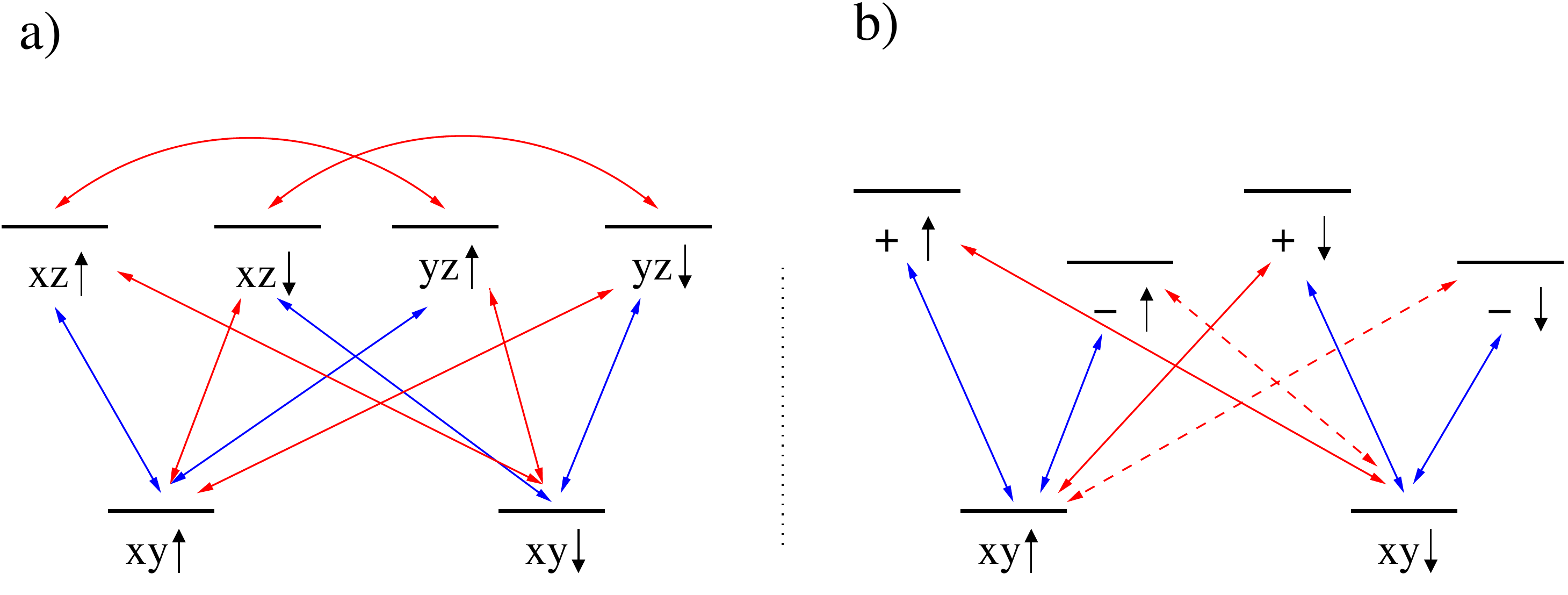}
\end{center}
\caption{Level structure and interactions of the three-band hamiltonian. Interactions
of $H_{aso}\sim\Delta_{aso}$ Eq.\,(\ref{eq:hatso}) are shown in
red and those of the asymmetry hopping $H_{I}$ Eq.\,(\ref{eq:ashop})
are indicated in blue. The red dashed lines correspond to a atomic
SO interaction which around the $\Gamma$-point (i.e. $\sim tk^{2}<\Delta_{aso}$)
is much smaller than the matrix elements represented by the solid
lines. Panel (a) corresponds to the original Hamiltonian whereas in
(b) the atomic SOC between $xz$ and $yz$ has been diagonalized.}
\label{fig5} 
\end{figure}

Around the $\Gamma$ point the non-interacting part of the Hamiltonian
(\ref{eq:hamiltonian0}) reads 
\begin{eqnarray}
\varepsilon_{k}^{xy} & = & (t_{1}+2t_{3})k^{2}\\
\varepsilon_{k}^{xz} & = & (t_{1}+t_{3})k_{x}^{2}+t_{2}k_{y}^{2}+\Delta\\
\varepsilon_{k}^{yz} & = & (t_{1}+t_{3})k_{y}^{2}+t_{2}k_{x}^{2}+\Delta\,,
\end{eqnarray}
where $k^{2}=k_{x}^{2}+k_{y}^{2}$. The atomic SOC {[}$\sim\tau^{z}$,
cf. Eq.\,(\ref{eq:hatso}){]} lifts the degeneracy between $xz$
and $yz$ but leaves the spin degeneracy, cf. Fig.\,\ref{fig5}.
One obtains the new  $-(+)$, corresponding to the pairs of bands (3,4) and (5,6) respectively, 
\begin{eqnarray*}
E_{\sigma}^{\pm} & = & \Delta+\frac{1}{2}(t_{1}+t_{2}+t_{3})k^{2}\\
 & \pm & \frac{1}{2}\sqrt{(t_{1}+t_{3}-t_{2})^{2}(k_{x}^{2}-k_{y}^{2})^{2}+4\Delta_{aso}^{2}}
\end{eqnarray*}
and eigenfunctions 
\begin{eqnarray}
|xz,\uparrow\rangle & = & \imath a|+,\uparrow\rangle+\imath b|-,\uparrow\rangle\label{traf1}\\
|yz,\uparrow\rangle & = & -b|+,\uparrow\rangle+a|-,\uparrow\rangle\label{traf2}\\
|xz,\downarrow\rangle & = & \imath a|+,\downarrow\rangle+\imath b|-,\downarrow\rangle\label{traf3}\\
|yz,\downarrow\rangle & = & b|+,\downarrow\rangle-a|-,\downarrow\rangle\label{traf4}
\end{eqnarray}
with 
\[
a\approx\frac{1}{\sqrt{2}}\sqrt{1+\frac{\varepsilon^{xz}-\varepsilon^{yz}}{(\varepsilon^{xz}-\varepsilon^{yz})^{2}+4\Delta_{aso}^{2}}}\approx\frac{1}{\sqrt{2}}\left(1+\frac{1}{4\Delta_{aso}}(t_{1}+t_{3}-t_{2})(k_{x}^{2}-k_{y}^{2})\right),\:\;b=\sqrt{1-a^{2}}
\]
In the basis $|+,\uparrow\rangle,|+,\downarrow\rangle,|-,\uparrow\rangle,|-,\downarrow\rangle$
the asymmetry hopping is given by 
\begin{eqnarray*}
H_{I} & = & \sqrt{2}\gamma(k_{y}+\imath k_{x})|xy,\uparrow\rangle\langle+,\uparrow|+h.c.\\
 & + & \sqrt{2}\gamma(k_{y}-\imath k_{x})|xy,\downarrow\rangle\langle+,\downarrow|+h.c.\\
 & + & \sqrt{2}\gamma(k_{y}-\imath k_{x})|xy,\uparrow\rangle\langle-,\uparrow|+h.c.\\
 & + & \sqrt{2}\gamma(k_{y}+\imath k_{x})|xy,\downarrow\rangle\langle-,\downarrow|+h.c.
\end{eqnarray*}
and the residual coupling of $H_{aso}$ with the $xy$-level reads
\begin{eqnarray*}
H_{aso} & = & \sqrt{2}\Delta_{aso}|xy,\uparrow\rangle\langle+,\downarrow|+h.c.\\
 & + & \sqrt{2}\Delta_{aso}|xy,\downarrow\rangle\langle+,\uparrow|+h.c.\\
 & - & \frac{1}{2\sqrt{2}}(t_{1}+t_{3}-t_{2})(k_{x}^{2}-k_{y}^{2})|xy,\uparrow\rangle\langle-,\downarrow|+h.c.\\
 & - & \frac{1}{2\sqrt{2}}(t_{1}+t_{3}-t_{2})(k_{x}^{2}-k_{y}^{2})|xy,\downarrow\rangle\langle-,\uparrow|+h.c.\,.
\end{eqnarray*}
In the following we restrict to the region close to the $\Gamma$
point, where $\Delta_{aso}>tk^{2}$, and neglect therefore the two
latter terms in $H_{aso}$ resulting in the effective coupling structure
depicted in panel (b) of Fig.\,\ref{fig5}. We can now calculate the effective
interactions between levels $\alpha$,$\beta$ in 2nd order perturbation
theory 
\begin{equation}
\langle\alpha|H^{(2)}|\beta\rangle=-\frac{1}{2}\sum_{n}\left(\frac{1}{E_{n}-E_{\alpha}}+\frac{1}{E_{n}-E_{\beta}}\right)H_{\alpha,n}H_{n,\beta}\label{eq:2ndorder}
\end{equation}
and $\alpha$,$\beta$ either corresponds to the $xy$ or to the $\pm$
levels. For the $xy$ states one finds 
\begin{eqnarray*}
\langle xy,\uparrow|H^{(2)}|xy,\downarrow\rangle & \approx & -\frac{\langle xy,\uparrow|H_{I}|+,\uparrow\rangle\langle+,\uparrow|H_{aso}|xy,\downarrow\rangle}{\Delta}\\
 & - & \frac{\langle xy,\uparrow|H_{aso}|+,\downarrow\rangle\langle+,\downarrow|H_{I}|xy,\downarrow\rangle}{\Delta}
\end{eqnarray*}
and similarly for $\langle xy,\downarrow|H^{2}|xy,\uparrow\rangle$.
Inserting the matrix elements yields an effective \textit{Rashba}
SOC $\sim k_{y}\tau^{x}-k_{x}\tau^{y}$ 
\begin{equation}
H_{xy}^{SOC}=-4\frac{\gamma\Delta_{aso}}{\Delta}\left(\begin{array}{cc}
0 & k_{y}+\imath k_{x}\\
k_{y}-\imath k_{x} & 0
\end{array}\right)=-\alpha_{xy}(\tau^{x}k_{y}-\tau^{y}k_{x})\label{eq:xyrashba}
\end{equation}
with a \textit{negative} coupling constant with $\alpha_{xy}=4\gamma\Delta_{aso}/\Delta$.

From Fig.\,\ref{fig5} one can see that the same matrix elements
also mediate the 2nd order interaction between the $+,\sigma$ and
$+,-\sigma$ states. Since in this case the denominator in Eq.\,(\ref{eq:2ndorder})
is negative we obtain a \textit{positive} coupling $\alpha_{+}=4\gamma\Delta_{aso}/\Delta$
for the $E^{+}$ states 
\begin{equation}
H_{E^{+}}^{SOC}=4\frac{\gamma\Delta_{aso}}{\Delta}\left(\begin{array}{cc}
0 & k_{y}-\imath k_{x}\\
k_{y}+\imath k_{x} & 0
\end{array}\right)=\alpha_{+}(\tau^{x}k_{y}+\tau^{y}k_{x})\,.\label{eq:eprashba}
\end{equation}
Moreover the off-diagonal matrix elements in Eq.\,(\ref{eq:eprashba})
are c.c. to those of Eq.\,(\ref{eq:xyrashba}) which means that the
$+,\sigma$ and $+,-\sigma$ states are interacting via a \textit{Dresselhaus}
coupling $\sim k_{y}\tau^{x}+k_{x}\tau^{y}$.

The effective interactions between $-,\sigma$ and $-,-\sigma$ can
be again obtained from 2nd order perturbation theory in the limit
$tk^{2}<\Delta_{aso}$ which now involves the matrix elements represented
by the dashed lines in Fig.\,\ref{fig5}. The resulting effective
coupling reads 
\begin{eqnarray}
H_{-}^{SOC} & = & -\frac{\gamma(t_{1}+t_{3}-t_{2})(k_{x}^{2}-k_{y}^{2})}{\Delta}\left(\begin{array}{cc}
0 & k_{y}+\imath k_{x}\\
k_{y}-\imath k_{x} & 0
\end{array}\right)=-\beta(k_{x}^{2}-k_{y}^{2})(\tau^{x}k_{y}-\tau^{y}k_{x})\label{eq:emrashba}
\end{eqnarray}
and therefore corresponds to a linear Rashba SOC but with a coupling
constant $\sim(k_{x}^{2}-k_{y}^{2})$.

\section{THE CLEAN LIMIT}

\label{sec:cleanlimit}

In this section we evaluate the Drude weight for the effective models
discussed in section \ref{sec:effectivemodels}.

\subsection{$xy$ bands}

In this case the eigenvalues and eigenvectors corresponding to the
Hamiltonian (\ref{eq:xyrashba}) read

\begin{equation}
E_{\pm}^{xy}=\frac{k^{2}}{2m}\pm\alpha_{xy}k,\;|\pm\rangle=\frac{1}{\sqrt{2}}\left(\begin{array}{c}
\mp\imath\mathrm{e}^{-\imath\theta}\\
1
\end{array}\right)\label{eq:Eigenxy}
\end{equation}
with $\hat{p}_{x}=\cos(\theta)$ and $\hat{p}_{y}=\sin(\theta)$.
As for the Rashba model (\ref{eq:Rashba2DEG}), the spin operator
is simply the Pauli matrix $S^{y}=\tau^{y}/2$ and the charge current
is similar to the 2DEG case $J_{x}=(-e)(k_{x}/m\tau^{0}+\alpha_{xy}\tau^{y}).$
The interband matrix elements read

\[
\langle\mathbf{k},s|\mathrm{S}^{y}|\mathbf{k},-s\rangle=\frac{1}{2}is\sin(\theta),\;\langle\mathbf{k},-s|\mathrm{J}_{x}|\mathbf{k},s\rangle=-is\sin(\theta)(-e)\alpha_{xy}
\]
and the response function (in the zero-temperature limit) gives

\[
R_{yx}^{'}(\omega\rightarrow0)=\frac{1}{2}(-e)\alpha_{xy}\sum_{\mathbf{p}\mathrm{s}}\sin^{2}(\theta)\frac{\,\theta(E_{s}^{xy}(\mathbf{p})-\mu)-\theta(E_{-s}^{xy}(\mathbf{p})-\mu)}{E_{s}^{xy}(\mathbf{p})-E_{-s}^{xy}(\mathbf{p})}=\frac{1}{2}(-e)\alpha_{xy}N_{0},
\]
which leads to the Drude weight

\begin{equation}
D_{xy}^{ISG}=\frac{\pi}{2}e\alpha_{xy}N_{0}\label{eq:Drudeweightxy}
\end{equation}
with an opposite sign as compared to the 2DEG case of Eq.\,(\ref{eq:Drude2DEG}).

\subsection{$E^{-}$ bands}

In this case the eigenvalues and eigenvectors corresponding to the
Hamiltonian (\ref{eq:emrashba}) read

\begin{equation}
E_{\pm}^{-}=\frac{k^{2}}{2m}\pm\beta|\zeta|\,k^{3},\;|\pm\rangle=\frac{1}{\sqrt{2}}\left(\begin{array}{c}
\mp\imath\frac{\zeta}{|\zeta|}\mathrm{e}^{-\imath\theta}\\
1
\end{array}\right)\label{eq:Eigenem}
\end{equation}
where $\zeta=\hat{k}_{x}^{2}-\hat{k}_{y}^{2}$ with $\hat{k}_{x}=\cos(\theta)$
and $\hat{k}_{y}=\sin(\theta)$. In this case the spin operator reads
$\mathrm{S}^{y}=-\gamma(k_{x}^{2}-k_{y}^{2})\tau^{y}/2$, while the
charge current has a more complicated structure as compared to the
2DEG case $\mathrm{J}_{x}=(-e)\left(\frac{k_{x}}{m}\tau^{0}-2\beta k_{x}k_{y}\tau^{x}+\beta(3k_{x}^{2}-k_{y}^{2})\tau^{y}\right)$.
The interband matrix elements read

\begin{eqnarray*}
\langle\mathbf{k},s|\tau^{y}|\mathbf{k},-s\rangle & = & \imath s\frac{\zeta}{|\zeta|}\sin(\theta)\\
\langle\mathbf{k},-s|\tau^{y}|\mathbf{k},s\rangle & = & -\imath s\frac{\zeta}{|\zeta|}\sin(\theta)\\
\langle\mathbf{k},s|\tau^{x}|\mathbf{k},-s\rangle & = & \imath s\frac{\zeta}{|\zeta|}\cos(\theta)\\
\langle\mathbf{k},-s|\tau^{x}|\mathbf{k},s\rangle & = & -\imath s\frac{\zeta}{|\zeta|}\cos(\theta).
\end{eqnarray*}
In the response function

\begin{eqnarray*}
R_{xy}^{'}(0) & = & -(-e)\sum_{\mathbf{k},,s}\frac{1}{2}(-\gamma k^{2})(\hat{k}_{x}^{2}-\hat{k}_{y}^{2})k^{2}\left[-2\beta\hat{k}_{x}^{2}\hat{k}_{y}^{2}+\beta(3\hat{k}_{x}^{2}\hat{k}_{y}^{2}-\hat{k}_{y}^{4})\right]\frac{\theta(\mu-E_{s}^{-}(\mathbf{k}))-\theta(\mu-E_{-s}^{-}(\mathbf{k}))}{2s\beta|\zeta|k^{3}}\\
 & = & (-e)\frac{\gamma}{2}\int_{0}^{2\pi}\frac{d\theta}{2\pi}(\hat{k}_{x}^{2}-\hat{k}_{y}^{2})^{2}\frac{\hat{k}_{y}^{2}}{|\zeta|}\int_{k_{-}}^{k_{+}}\frac{dk}{2\pi}k^{2}\\
 & = & (-e)\frac{\gamma}{2}\int_{0}^{2\pi}\frac{d\theta}{2\pi}(\hat{k}_{x}^{2}-\hat{k}_{y}^{2})^{2}\frac{\hat{k}_{y}^{2}}{|\zeta|}\frac{k_{+}^{3}-k_{-}^{3}}{6\pi}\\
 & = & (-e)\gamma k_{F}^{2}\beta k_{F}^{2}N_{0}\left(-\frac{1}{4}\right)
\end{eqnarray*}
the factors $\zeta$ disappear and the Drude weight reads

\begin{equation}
D_{-}^{ISG}=(-e)\left(\frac{\pi}{4}\gamma p_{F}^{2}\beta p_{F}^{2}N_{0}\right).\label{eq:Drudeem}
\end{equation}

\subsection{$E^{+}$ bands}

In this case the eigenvalues and eigenvectors corresponding to the
Hamiltonian (\ref{eq:eprashba}) read

\begin{equation}
E_{\pm}^{xy}=\frac{k^{2}}{2m}\pm\alpha_{+}k,\;|\pm\rangle=\frac{1}{\sqrt{2}}\left(\begin{array}{c}
\mp\imath\mathrm{e}^{\imath\theta}\\
1
\end{array}\right)\label{eq:Eigenep}
\end{equation}
with $\hat{k}_{x}=\cos(\theta)$ and $\hat{k}_{y}=\sin(\theta)$.
In this case the spin operator reads $\mathrm{S}^{y}=\gamma(k_{x}^{2}-k_{y}^{2})\tau^{y}/2$,
while the charge current is $\mathrm{J}_{x}=\left(-e\right)\left(\frac{p_{x}}{m}\tau^{0}+\alpha_{+}\tau^{y}\right)$.
The interband matrix elements are

\begin{eqnarray*}
\langle\mathbf{k},s|\tau^{y}|\mathbf{k},-s\rangle & = & -\imath s\sin(\theta),\\
\langle\mathbf{k},-s|\tau^{y}|\mathbf{k},s\rangle & = & \imath s\sin(\theta).
\end{eqnarray*}
The response function is

\begin{eqnarray*}
R_{xy}^{'}(0) & = & -(-e)\gamma\sum_{\mathbf{k},s}k^{2}(\hat{k}_{x}^{2}-\hat{k}_{y}^{2})\hat{k}_{y}^{2}\frac{\alpha_{+}}{2}\frac{\theta(\mu-E_{s}^{+}(\mathbf{k}))-\theta(\mu-E_{-s}^{+}(\mathbf{k}))}{2s\alpha_{+}k}\\
 & = & -(-e)\frac{\gamma}{2}\int_{0}^{2\pi}\frac{d\theta}{2\pi}(\hat{k}_{x}^{2}-\hat{k}_{y}^{2})\hat{k}_{y}^{2}\int_{k_{-}}^{k_{+}}\frac{k^{2}dk}{2\pi}\\
 & = & (-e)\gamma k_{F}^{2}\alpha_{+}N_{0}\left(-\frac{1}{4}\right)
\end{eqnarray*}

and the Drude weight is

\begin{equation}
D_{+}^{ISG}=(-e)\left(\frac{\pi}{4}\gamma p_{F}^{2}\alpha_{+}N_{0}\right).\label{eq:Drudeep}
\end{equation}

\section{THE DISORDERED LIMIT}

\label{sec:disorderedlimit}

It is well known that in the presence of disorder, the Drude weight
in the formula for the optical conductivity is suppressed and the
spectral weight goes into the regular part. In the Drude model, the
regular part, as function of the frequency, has a Lorentzian shape whose
width is controlled by the scattering rate $\tau^{-1}$ (not to be
confused with the Pauli matrices). Such a transfer of spectral weight
from the singular to the regular part occurs also in the case of the
SGC. To this end we need to introduce disorder in our model. This
will be done in the numerical computation of the next section, whereas
in this section we introduce disorder within the effective models
derived in section \ref{sec:effectivemodels} by using the standard
diagrammatic impurity technique. This technique has been applied to
the Rashba model for the evaluation of the ISG effect\cite{Edelstein1990},
anisotropy magnetoresistance\cite{Raimondi2001,Schwab2002} and spin
Hall effect\cite{Raimondi2005}. We review here the basic aspects
by focusing on the case of the xy-bands, which is equivalent to the
Bychkov-Rashba model in the 2DEG. By following the standard procedure,
disorder is introduced as a random potential $V(\mathbf{r})$, with
zero average $\left\langle V(\mathbf{r})\right\rangle =0$ and white-noise
correlations $\left\langle V(\mathbf{r})V(\mathbf{r})\right\rangle =n_{i}u^{2}\delta(\mathbf{r}-\mathbf{r}')$,
with $n_{i}$ being the impurity concentration. By Fermi golden rule,
one associates a scattering rate $\tau^{-1}=2\pi n_{i}u^{2}N_{0}$,
where $N_{0}$ is the single-particle density of state previously
introduced in Eq.\,(\ref{eq:Drude2DEG}). We will consider the weak-disorder
limit which is controlled by the small parameter $(E_F\tau)^{-1}$,
with $E_{F}$ the Fermi energy. In the diagrammatic impurity technique,
the first step is the introduction of the irreducible self-energy
in the self-consistent Born approximation for the electron Green function.

\subsection{The case of the $E^{xy}$ bands }

\begin{figure}
\begin{center}
\includegraphics{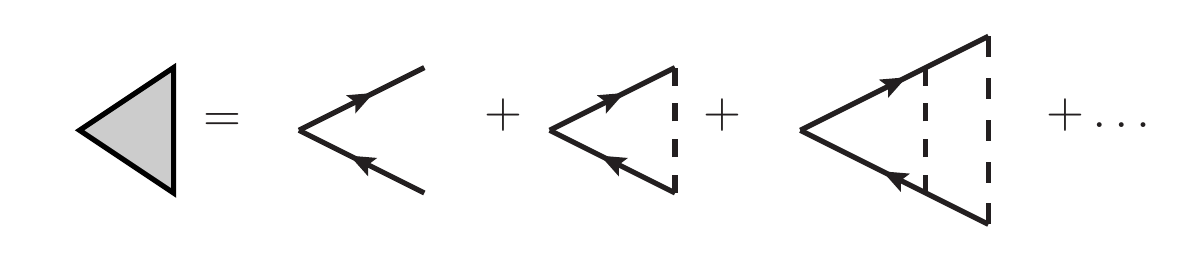}
\end{center}
\caption{Ladder diagrams for the determination of the dressed vertex. The gray-filled
triangle represents the infinite sum of diagrams, which results from
repeated scattering. The solid lines with arrows are Green function
propagators for electrons, whereas the dashed lines represent the
operation of impurity average.\label{fig:Ladder_diagrams} }

\end{figure}

The Green function, due to the SOC of the lowest pair of bands $H_{xy}^{SOC}$
of Eq.\,(\ref{eq:xyrashba}), can be expanded in Pauli matrices as
$\mathrm{G}=G_{0}\tau^{0}+G_{1}\tau^{x}+G_{2}\tau^{y}$ and explictly
reads

\begin{equation}
\mathrm{G}(\epsilon,\mathbf{k})=\frac{G_{+}(\epsilon,\mathbf{k})+G_{-}(\epsilon,\mathbf{k})}{2}\tau^{0}-(\tau^{x}\hat{k}_{y}-\tau^{y}\hat{k}_{x})\frac{G_{+}(\epsilon,\mathbf{k})-G_{-}(\epsilon,\mathbf{k})}{2}\label{eq:PauliGreenFunction}
\end{equation}
where

\begin{equation}
G_{\pm}(\epsilon,\mathbf{k})=\left[\epsilon-E_{\pm}^{xy}(\mathbf{k})-\Sigma(\epsilon)\right]^{-1},\label{eq:HelicityGF}
\end{equation}
and the self-energy has the form

\begin{equation}
\Sigma(\epsilon)=\mp\frac{\imath}{2\tau}\tau^{0},\label{eq:BornSE}
\end{equation}
the minus and plus signs applying to the retarded (R) and advanced
(A) sectors, respectively. The scattering time $\tau$ entering Eq.\,(\ref{eq:BornSE})
is exactly the one required by the Fermi golden rule. It is worth
noticing that the self-energy is proportional to the identity matrix
in the spin space\cite{Schwab2002}. Once the Green function is known, we may compute
the SGC by means of the Kubo formula

\begin{equation}
\sigma^{ISG}=\frac{1}{2\pi}\left\langle \mathrm{Tr}\left[\mathrm{S}^{y}\mathrm{G}^{R}\mathrm{J}_{x}\mathrm{G}^{A}\right]\right\rangle _{\mathrm{dis\,av}}\label{eq:KuboGF}
\end{equation}
which can be obtained from the expression (\ref{eq:LinRes}), after
averaging over the disorder configurations, represented as $\left\langle \ldots\right\rangle _{\mathrm{dis\,av}}$.
In the above the $\mathrm{Tr\ldots}$ symbol involves all degrees
of freedom, i.e. spin and space coordinates. The disorder average
in Eq.\,(\ref{eq:KuboGF}) enters in two ways. The first is to use
the disorder-averaged Green function given in Eq.\,(\ref{eq:PauliGreenFunction}).
The second is the introduction, to lowest order in the expansion parameter
$(E_{F}\tau)^{-1}$, of the so-called ladder diagrams, which lead
to vertex corrections. The vertex corrections procedure can be performed
either for the spin or charge vertex of Eq.\,(\ref{eq:KuboGF}).
Here we consider the vertex correction for the charge current vertex.
The \emph{dressed} vertex $\tilde{\mathrm{J}}_{x}$ obeys the Bethe--Salpeter
equation

\begin{equation}
\tilde{\mathrm{J}}_{x}=\mathrm{J}_{x}+n_{i}u^{2}\sum_{\mathbf{k}}\mathrm{G}^{R}(\epsilon,\mathbf{k})\tilde{\mathrm{J}}_{x}\mathrm{G}^{A}(\epsilon,\mathbf{k}),\label{eq:vertexequation}
\end{equation}
which results from the infinite summation of ladder diagrams, as shown
in Fig. \ref{fig:Ladder_diagrams}. In terms of the dressed vertex
the SGC reads

\begin{equation}
\sigma^{ISG}=\frac{1}{2\pi}\sum_{\mathbf{k}}\mathrm{tr}\left[\mathrm{S}^{y}\mathrm{G}^{R}(\epsilon,\mathbf{k})\tilde{\mathrm{J}}_{x}\mathrm{G}^{A}(\epsilon,\mathbf{k})\right],\label{eq:KubAv}
\end{equation}
where now the lower case trace symbol involves the spin degrees of
freedom only. The problem is then reduced to the solution of the Bethe--Salpeter
equation (\ref{eq:vertexequation}) and to the evaluation of the bubble
(\ref{eq:KubAv}). In general the Bethe--Salpeter equation is an
integral equation. However, in the present case of white-noise disorder,
the Bethe--Salpeter equation becomes an algebraic one, even though
still having a spin structure. In the appendix \ref{sec:BSVertex} we
provide the details of the solution of Eq.\,(\ref{eq:vertexequation}),
which leads to

\begin{equation}
\tilde{\mathrm{J}}_{x}=(-e)\frac{k_{x}}{m},\label{eq:dressedvertexxy}
\end{equation}
which shows that the vertex corrections exactly cancel the interband
matrix elements of the charge current vertex. As a result, the evaluation 
of Eq.\,(\ref{eq:KubAv}) leads to

\begin{equation}
\sigma_{xy}^{ISG}=eN_{0}\alpha_{xy}\tau,\label{eq:Edelsteinxy}
\end{equation}
which must be compared with the Drude weight evaluated in Eq.\,(\ref{eq:Drudeweightxy}).

\subsection{The case of the $E^{-}$ bands}

According to the analysis of appendix \ref{sec:BSVertex}, the dressed
charge current vertex reads

\[
\tilde{\mathrm{J}}_{x}=\left(-e\right)\left[\frac{k_{x}}{m}\tau^{0}-2\beta k_{x}k_{y}\tau^{x}+\beta(3k_{x}^{2}-k_{y}^{2})\tau^{y}+\frac{1}{4}\beta p_{F}^{2}\tau^{y}\right].
\]
The evaluation then of Eq.\,(\ref{eq:KubAv}) leads to

\begin{equation}
\sigma_{-}^{ISG}=-\frac{5}{8}eN_{0}\gamma\beta p_{F}^{4}\tau,\label{eq:Edelsteinm}
\end{equation}
which has a sign opposite to that of the $E^{xy}$ bands. In Eq.\,(\ref{eq:Edelsteinm})
the combination $\beta p_{F}^{2}$ plays the role of an effective
SOC, whereas $\gamma p_{F}^{2}$ is the spin dressing factor accounting
for the interactions in the original model.

\subsection{The case of the $E^{+}$ bands}

According to the analysis of appendix \ref{sec:BSVertex}, the dressed
charge current vertex reads

\[
\tilde{\mathrm{J}}_{x}=(-e)\frac{k_{x}}{m}\tau^{0}.
\]
The evaluation then of Eq.\,(\ref{eq:KubAv}) leads to

\begin{equation}
\sigma_{+}^{ISG}=(\gamma p_{F}^{2})eN_{0}\alpha_{+}\tau,\label{eq:Edelsteinp}
\end{equation}
which shows again a change of sign with respect to that of the $E^{-}$
bands. Also here the combination $\gamma p_{F}^{2}$ is the spin dressing
factor accounting for the interactions in the original model.

\section{THE NUMERICAL APPROACH}

In this section we present our numerical results. The starting point
is the response function defined in Eq.\,(\ref{eq:LinRes}), which
may be expressed as follows

\begin{equation}
R_{yx}(\omega)=\frac{1}{N}\sum_{k,p}(f_{p}-f_{k})\frac{\langle p|\mathrm{S}^{y}|k\rangle\langle k|\mathrm{J}_{x}|p\rangle}{\omega+i\eta+E_{p}-E_{k}},\label{eq:LinResNum}
\end{equation}
where $k$ and $p$ are quantum numbers labelling the eigenstates
of the Hamiltonian. For instance, in the absence of disorder, the
index $k$ includes the crystal momentum, the orbital and spin degrees
of freedom. The symbol $f_{k}$ stands for the Fermi function evaluated
at the energy of the eigenstate $k$. In Eq.\,(\ref{eq:LinResNum})
$N$ is the number of lattice sites. The numerical evaluation is performed
on a finite system and then it is convenient to separate from the
outset the Drude singular weight from the regular part as follows

\begin{equation}
D^{ISG}=-\frac{\pi}{N}\sum_{k,p}\frac{f_{p}-f_{k}}{E_{p}-E_{k}}\Re\langle p|\mathrm{S}^{y}|k\rangle\langle k|\mathrm{J}_{x}|p\rangle,\label{eq:DrudeNum}
\end{equation}

\begin{equation}
\sigma_{reg}^{ISG}=\lim_{\omega\rightarrow0}\mathcal{P\frac{\mathrm{\mathit{R_{yx}^{''}(\omega)}}}{\omega}=}-\frac{1}{N}\sum_{k,p}\frac{f_{p}-f_{k}}{E_{p}-E_{k}}\Im\frac{\langle p|S^{y}|k\rangle\langle k|J_{x}|p\rangle}{i\eta+E_{p}-E_{k}},\label{eq:RegNum}
\end{equation}
where $\Re$ and $\Im$ indicate the real and imaginary parts. 

\begin{figure}
\begin{center}
\includegraphics[width=7cm]{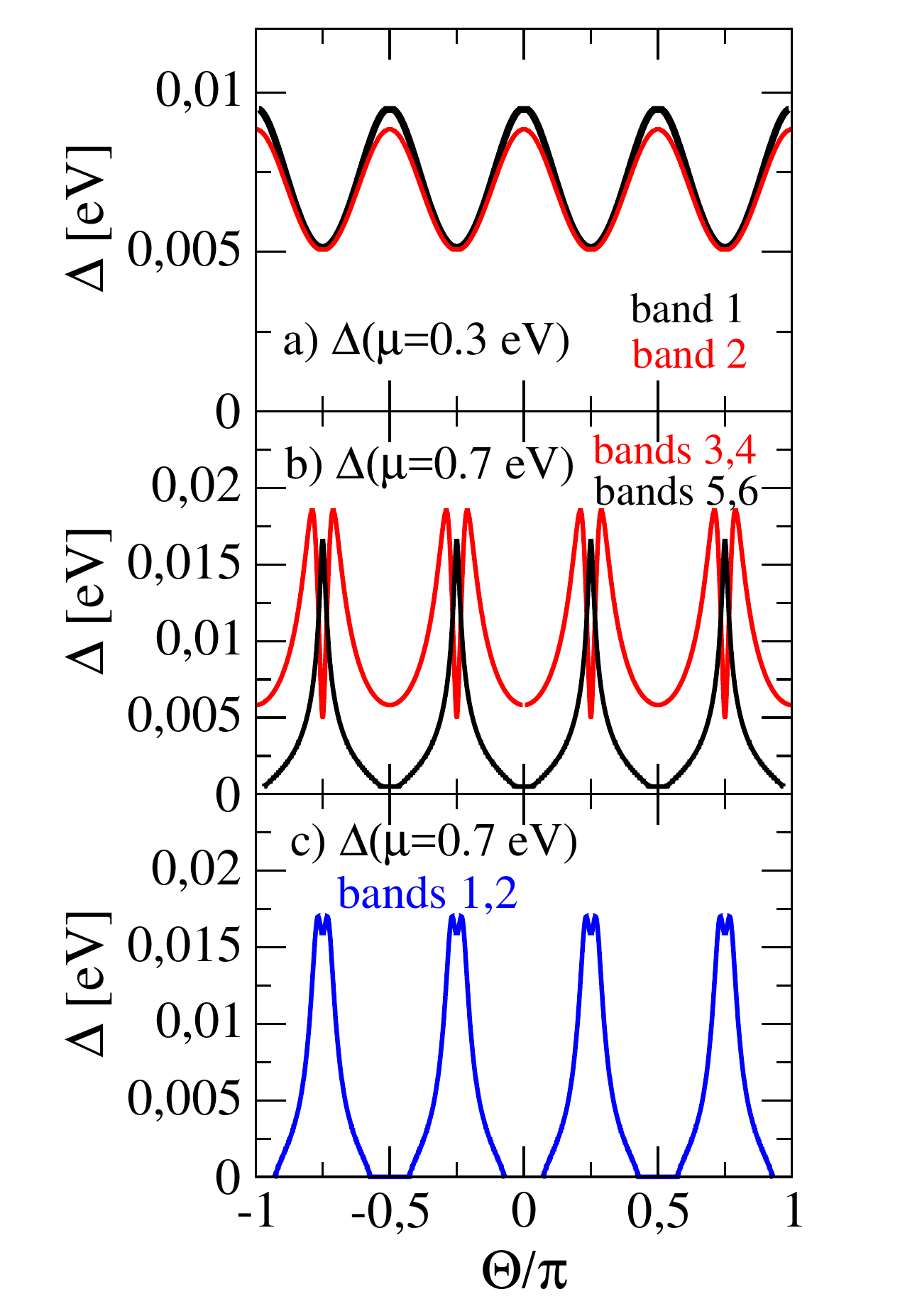}
\end{center}
\caption{Size of the spin-orbit split gap around the Fermi surface for $\mu=0.3$
eV (a) and $\mu=0.7$\,eV (b,c). This is obtained by determining
the cut $k_{F}$ of each band with $\mu$ and then calculating the
energy difference to the 'other' SO split band at the same $k_{F}$.
If the SO interaction has some significant momentum dependence around
$k_{F}$ the gap determined for each of the two bands from a pair
may slightly differ as in panel (a). The angle $\Theta$ is defined
with respect to the $k_{x}$-axis.\label{fig:figapp}}
\end{figure}

Fig.\,\ref{fig:figapp} shows the behavior of the spin-orbit split
gap at the Fermi surface. Inspection of Fig.\,\ref{fig:figapp} reveals
that for $\mu=0.3$\,eV the gap has extrema at energies $\Delta\approx0.005$
and $0.009$\,eV which are expected to dominate the response due
to the ``saddle-point'' character of the corresponding states as
discussed below. For $\mu=0.7$ , as shown in Fig.\,\ref{fig:chiralities},
all bands are occupied and all the gaps will appear in the response
function. In particular the pair of bands (1,2) contributes to the
gap at energies $\Delta\approx0.015$ eV, the pair of bands (3,4)
at energies $\Delta\approx0.005-0.015$ eV, and the pair of bands
(5,6) at energies $\Delta\approx0$.

\begin{figure}[H]
\begin{center}
\includegraphics[width=8cm]{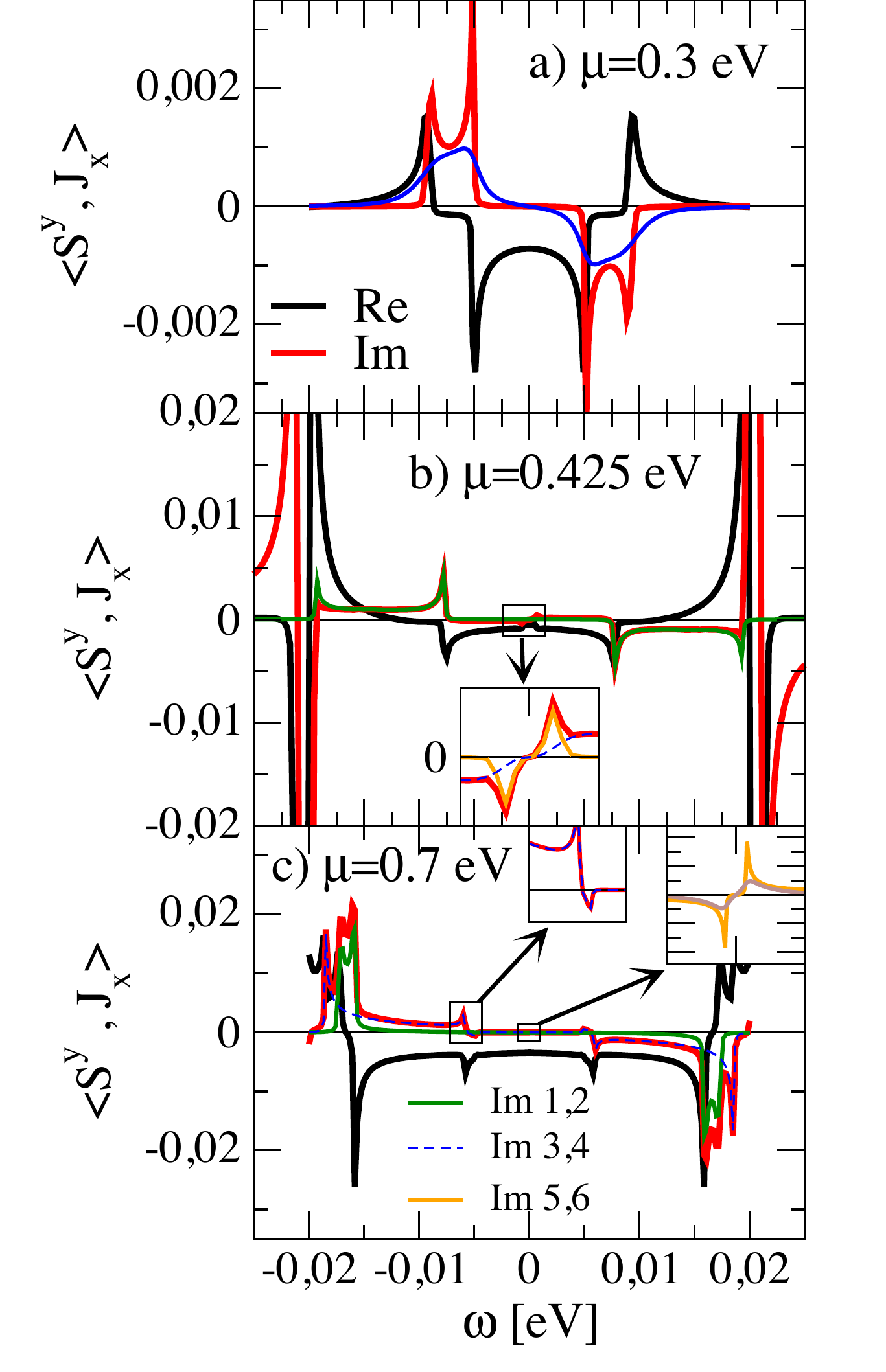}
\end{center}
\caption{Frequency dependent real (black line) and imaginary part (red line)
of the spin-current correlation function $R_{yx}(\omega)$ evaluated
for chemical potentials $\mu=0.3$\,eV (a), $\mu=0.425$ eV (b) and
$\mu=0.7$\,eV (c) and lifetime parameter $\eta=5\cdot10^{-5}$\,eV
. In panel (a), the additional blue line is for $\eta=10^{-3}$ eV
and the slope of the imaginary part at $\omega=0$ defines the SGC
$\sigma_{reg}^{ISG}$. In panel (b), the individual contribution
of the pair of bands (1,2) (green line), (3,4) (blue dashed line)
and (5,6) (yellow line) is also shown. The inset details the behavior
around $\omega=0$, dominated by the pairs of bands (3,4) and (5,6).
In panel (c) the individual contribution of the different pairs of
bands is shown as in panel (b). The inset around $\omega=0$ evidences
the contribution from the pair (5,6) at two different values of the
lifetime parameter $\eta=10^{-4}$(brown line) and $\eta=10^{-5}$(yellow
line), whereas the inset around $\omega=0.075$ shows the contribution
from the pair of bands (3,4) at $\eta=10^{-4}$ (blue dashed line)
and $\eta=10^{-5}$(magenta line).\label{fig:fig_sISG}}
\end{figure}

Fig.\,\ref{fig:fig_sISG} shows the frequency dependence of the real
and imaginary parts of the response function $R_{yx}(\omega)$ for
the three chemical potentials $\mu=0.3$\,eV, $\mu=0.425$ eV and
$\mu=0.7$\,eV.  The underlying ground state is for a
homogeneous system but we investigate the influence of the particle-hole
lifetime parameter $\eta$. As compared with the approach discussed in
Sec. \ref{sec:disorderedlimit} this mimics the inclusion of momentum relaxation
without considering vertex corrections. 
For $\mu=0.3$ eV, when only the lowest pair of bands
is occupied, one may interpret the observed behavior in terms of the
Rashba model of Eq.\,(\ref{eq:Rashba2DEG}). The low energy structure
is determined by transitions across the spin-orbit split gap of the
same $t_{2g}$ band. In fact, it is exactly in this energy range that
the imaginary part of $R_{yx}(\omega)$ develops a peak structure
as discussed in Eq.(\ref{eq:imaginaryResponse}). In the clean system,
i.e. lifetime parameter $\eta\rightarrow0$, the imaginary part vanishes
for energies below the minimum gap excitations and therefore the slope
of $R_{yx}^{''}(\omega)$, which determines the regular ISG response
$\sigma_{yx,reg}^{ISG}$, is zero, whereas the limiting value of the
real part fixes the Drude weight of the singular contribution. As
shown in Fig.\,\ref{fig:fig_sISG} (panel (a)) a finite $\eta$ (or,
similarly, a finite temperature) broadens the excitations and therefore
induces a finite slope of $R_{yx}^{''}(\omega)$ at $\omega=0$ leading
thus to a finite ISG response. This is evidenced by making $\eta$
larger in Fig.\,\ref{fig:fig_sISG}: the red curve is for $\eta=10^{-5}$,
while the blue one for $\eta=10^{-3}$. The fact, that the ISG response
vanishes for a clean system is consistent with the analysis in Ref.
\cite{Shen2014} and with the discussion at the end of section \ref{sec:LRT}.
According to the KKR (\ref{eq:KK}), the sign of the imaginary part
of the ISG response function is determined by the sign of the effective
Rashba SOC. The negative sign shown by the numerical evaluation of
Fig.\,\ref{fig:fig_sISG} agrees with the sign found for the coupling
$-\alpha_{xy}$ in the effective model for the lowest pair of $xy$
bands in Eq.\,(\ref{eq:Drudeweightxy}). We remind that the Drude
weight is determined by the zero-frequency value of the real part,
which is obtained from the imaginary part via the KKR (\ref{eq:KK}).

For the chemical potential $\mu=0.425$ eV, close to the Lifshitz
point, all the bands are occupied and the imaginary part of the response
function gets contributions from the interband transitions across
the spin-orbit split gaps of all the pairs of bands as well as from
the interband transitions involving different pairs of bands simultaneously.
In panel (b) of Fig.\,\ref{fig:fig_sISG} this is evidenced by showing,
together with the full imaginary part (red line) also the contribution
of the individual pairs of bands: (1,2) (green line), (3,4) (blue
dashed line) and (5,6) (yellow line). From this we conclude that the
large spectral weight at energy $\omega=0.02$ is due to interband
transition between different pairs of bands. The green curve shows
that the contribution due exclusively to the lowest pair of bands
(1,2) is still around the same energy as in panel (a) and hence the
behavior of this pair of bands is still well described by the effective
Rashba model of Eq.(\ref{eq:xyrashba}). On the other hand, the inset
around zero energy in panel (b) shows how the low energy contribution
is dominated by both the pairs of bands (3,4) and (5,6), which at
this chemical potential have a small Fermi surface and a small spin-orbit
split gap. Notice that the sign of the imaginary part is opposite
to that of bands (1,2), indicating an opposite sign for the Drude
spectral weight in the limit of vanishing lifetime parameter $\eta$
in agreement with Eqs.(\ref{eq:Drudeem}) and (\ref{eq:Drudeep}).
Furthermore one may notice that for the small but finite value used
for the lifetime parameter $\eta$ both pairs of bands yield a finite
positive slope at zero frequency. Whereas the regular part at zero
frequency has the sign due to the pair or pairs of bands with the
lowest gap, the zero-frequency value of the real part, which is associated
to the Drude spectral weight, is obtained from the integrated spectral
weight of all the interband transitions. As a consequence, also interband
transitions at high energy may contribute provided they have a strong
spectral weight, which must compensate the big frequency denominator
of the KKR relation (\ref{eq:KK}). As it is apparent from panel (b),
close to the Lifshitz point, the very small value of the gap of the
pairs of bands (3,4) and (5,6) is sufficient to determine a positive
value of the zero-frequency real part.

For the chemical potential $\mu=0.7$ eV as well, the energy response
is determined by the gap structure of all the bands (1,2), (3,4) and
(5,6), as is evident from Fig.\,\ref{fig:chiralities}, even though
now we are far away from the Lifshitz point. As shown in panel (c)
of Fig.\,\ref{fig:fig_sISG} and the inset at zero frequency, the
peak coming from the smallest gap excitations at $\omega\approx0.0001$\,eV
belongs to the pair of bands (5,6) with the smaller $k_{F}$. The
next higher excitation comes from the pair of bands (3,4). As a result
at a finite value of the lifetime parameter, the regular part of the
ISG response is finite and positive. However, in this case, in contrast
to what happens close to the Lifshitz point, the opposite-in-sign
spectral weight of the interband transitions at higher energies is
sufficiently strong to drive the sign of the real part to a negative
value.

\begin{figure}[H]
\begin{center}
\includegraphics[width=8cm]{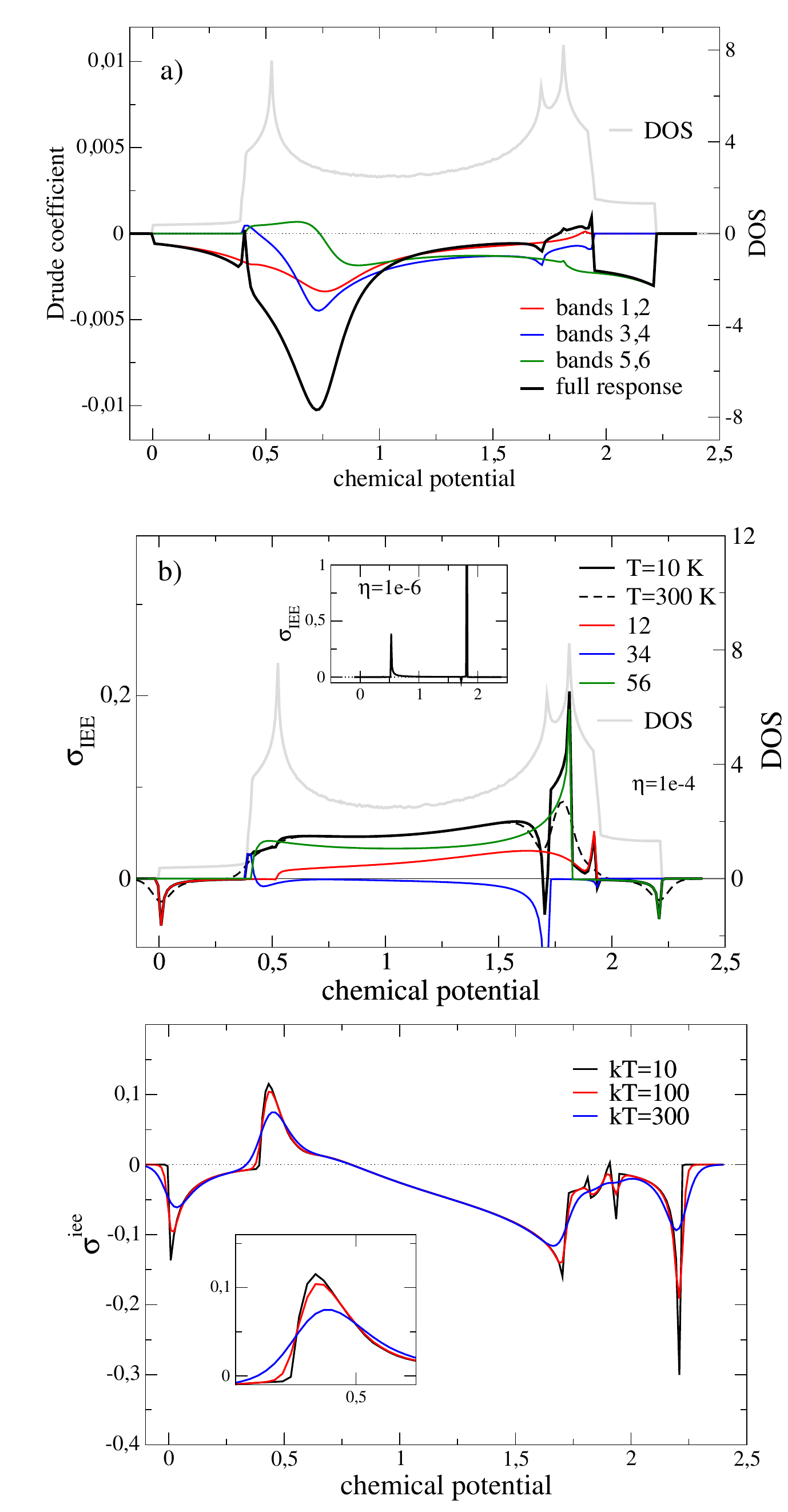}
\end{center}
\caption{Top panel: Drude coefficient at $T=10$\,K (black) of the ISG response
and the contribution of the individual bands. Middle panel: Regular
part of the inverse ISG response as a function of chemical potential
for lifetime parameter $\eta=1\cdot10^{-4}$ and the contribution
of the individual bands. In both panels the DOS is shown in grey for
comparison. The inset to the middle panel reports the regular part
for $\eta=1\cdot10^{-6}$ and $T=10$. Lowest panel: Regular part
for $\eta=1\cdot10^{-3}$ and different temperatures. The inset to
the lowest panel resolves the region around the Lifshitz point with
a significant temperature dependence. Calculations have been done
for a lattice with $6354\times6354$ k points.\label{fig:DrudekT10}}
\end{figure}

The analysis carried out in Fig. \,\ref{fig:fig_sISG} can be extended
to all values of the chemical potential and the result is reported
in Fig. \,\ref{fig:DrudekT10}. Panel (a) of Fig.\,\ref{fig:DrudekT10}
shows the full Drude spectral weight together with the contribution
of the individual pairs of bands at $T=10$\,K as a function of the
chemical potential. The Drude part, which is associated to the integrated
imaginary part of the response, does not depend significantly on the
temperature and on the lifetime parameter $\eta$ (Fig.\,\ref{fig:DrudekT10}
is for $\eta=1\cdot10^{-6}$ ). Close to the $\varGamma$ points of
all the bands one finds a negative Drude coefficient for the pair
of bands (1,2) and a positive coefficient for the pairs of bands (3,4)
and (5,6). This is in agreement with results of the effective model
discussed in section \ref{sec:effectivemodels}. 

Panel (b) of Fig.\,\ref{fig:DrudekT10} reports on the other hand
the regular part of the ISG response as function of the chemical potential.
For small lifetime parameter $\eta=1\cdot10^{-6}$ (inset) the response
is only significant around the energies where the DOS displays a van-Hove
singularity. In particular, the response at low chemical potentials
is suppressed because there the spin-orbit split gap is large and
$\eta=1\cdot10^{-6}$ is not sufficient to broaden the excitations
up to $\omega=0$. On the other hand, for $\eta=1\cdot10^{-4}$ (main
panel) one now observes a ISG response at all energies and also the
sign change upon crossing around the Lifhitz point as discussed for
panel (b) of Fig. \,\ref{fig:fig_sISG}. Such a sign change has been
also found in the experiment of Ref. \cite{Lesne2016}.

It can also be seen that the total ISG regular response is given by
the sum of the three contributions coming from the interband transitions
between each of the three pairs of the spin-orbit split bands. In
fact, we have seen that a finite $\sigma^{ISG}$ requires a broadening
of the same order than the energy of the contributing low energy excitation.
Therefore interband transitions between different pairs of bands cannot
 contribute due to their high excitation energies.

Since we investigate a clean system, we also obtain a finite value
for the Drude part $D^{ISG}$ which we checked not to depend on the
system size but is a robust result. In the presence of (real) disorder
we expect $D^{ISG}=0$ which then guarantees the stationarity of the
solution.

\begin{figure}
\begin{center}
\includegraphics[width=8cm]{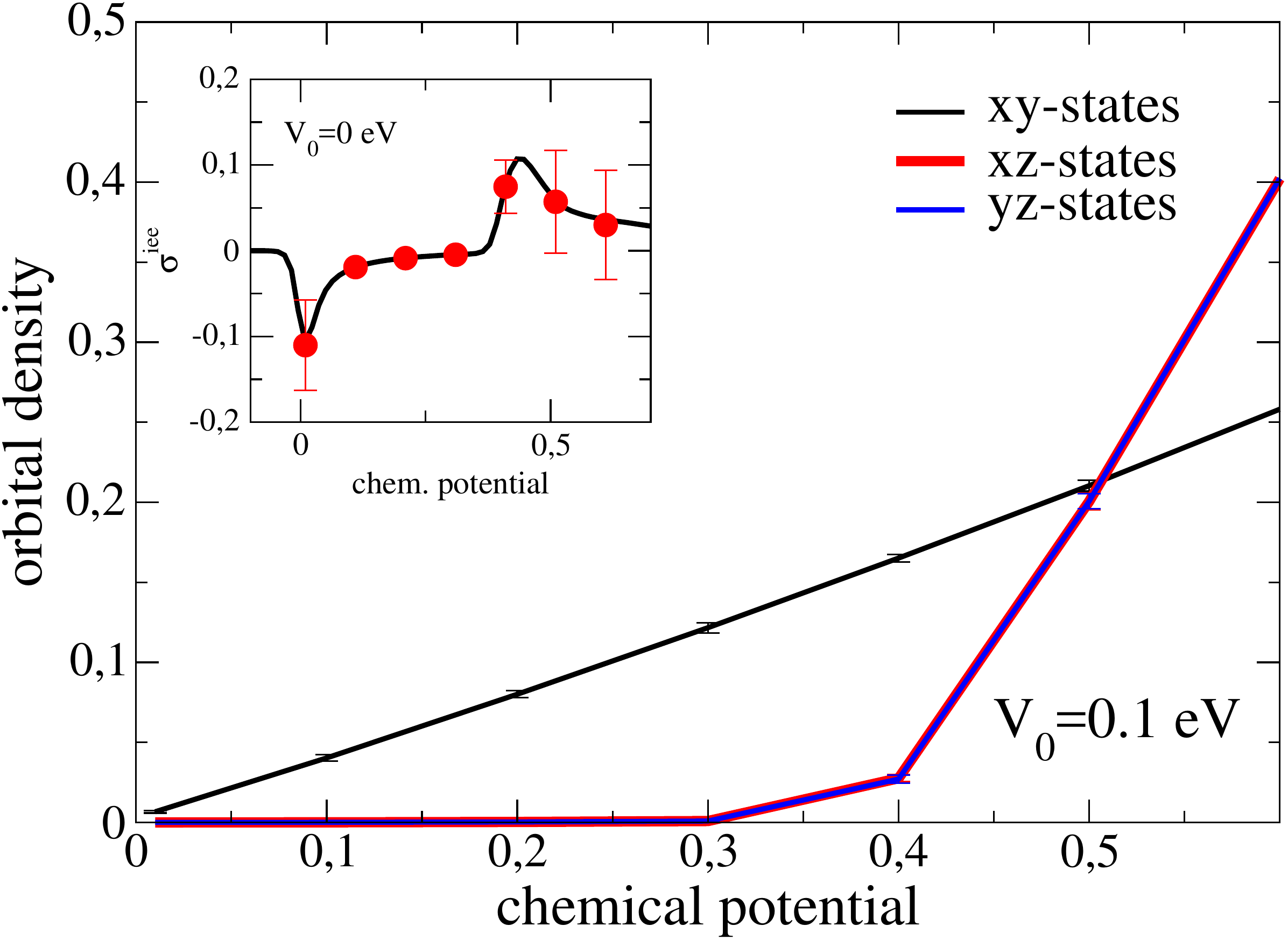}
\end{center}
\caption{Inset: Comparison of the ISG response for a homogeneous ($V_{0}=0$)
$24\times24$ lattice (red symbols) with the calculation for a $3762\times3762$
lattice (black line). Main panel: Averaged orbital occupation as a
function of chemical potential for a $24\times24$ lattice and disorder
potential $V_{0}=0.1$\,eV. Further parameters: temperature $T=100$
K, $\eta=1\cdot10^{-3}$\,eV.\label{fig:densfinite}}
\end{figure}

To implement the effect of disorder scattering we perform the calculation
of the SGC on finite lattices. In order to reduce the finite size
effects we average over twisted boundary conditions, i.e. for a $L_{x}\times L_{y}$
lattice we set

\[
|\Psi(R_{i})\rangle=\mathrm{e}^{\imath\Phi_{x,y}}|\Psi(R_{i}+L_{x,y})\rangle
\]
with $\Phi_{x,y}\in[0,2\pi]$ and we typically average over $50$
randomly chosen $(\Phi_{x},\Phi_{y})$. The inset to Fig.\,\ref{fig:densfinite}
demonstrates that the averaged finite lattice computation reproduces
the doping dependent SGC of the 'infinite' lattice calculation.

Disorder is introduced by a random local potential 
\[
\hat{V}=\sum_{i,\sigma}V_{i}\left(|xy_{i,\sigma}\rangle\langle xy_{i,\sigma}|+|xz_{i,\sigma}\rangle\langle xz_{i,\sigma}|+|yz_{i,\sigma}\rangle\langle yz_{i,\sigma}|\right)
\]
with $V_{i}$ randomly chosen on each site in the interval $[-V_{0},+V_{0}]$.
We then compute the SGC at some specified values of the chemical potential
and average over phases $\Phi_{x,y}$ and the disorder configurations.
The main panel of Fig.\,\ref{fig:densfinite} demonstrates that for
$V_{0}=0.1$\,eV the averaged orbital occupations are still well
defined for a given value of the chemical potential.

According to the analytical results of section \ref{sec:disorderedlimit},
in the presence of disorder, the SGC is positive for the pair of bands
(1,2) due to the $d_{xy}$ orbitals (cf. Eq.(\ref{eq:Edelsteinxy})),
is negative for the pair of bands (3,4) (cf. Eq.(\ref{eq:Edelsteinm})),
associated to the effective model of bands $E^{-},$ finally is positive
again for the pair of bands (5,6) (cf. Eq.(\ref{eq:Edelsteinp})),
associated with bands $E^{+}$. One then would expect a double change
of sign as the chemical potential enters the bottom of the different
pairs of bands. The numerical analysis of the clean limit with inclusion
of the effect of all the bands has shown a more complex behavior.
Close to the $\Gamma$ point, the behavior of the regular SG response
at zero frequency of the individual bands is well described by the
effective model. Instead, the Drude weight, which also includes all
interband transitions, cannot be simply interpreted in terms of the
individual contributions of the different pairs of bands. 

\begin{figure}
\begin{center}
\includegraphics[width=0.5\columnwidth]{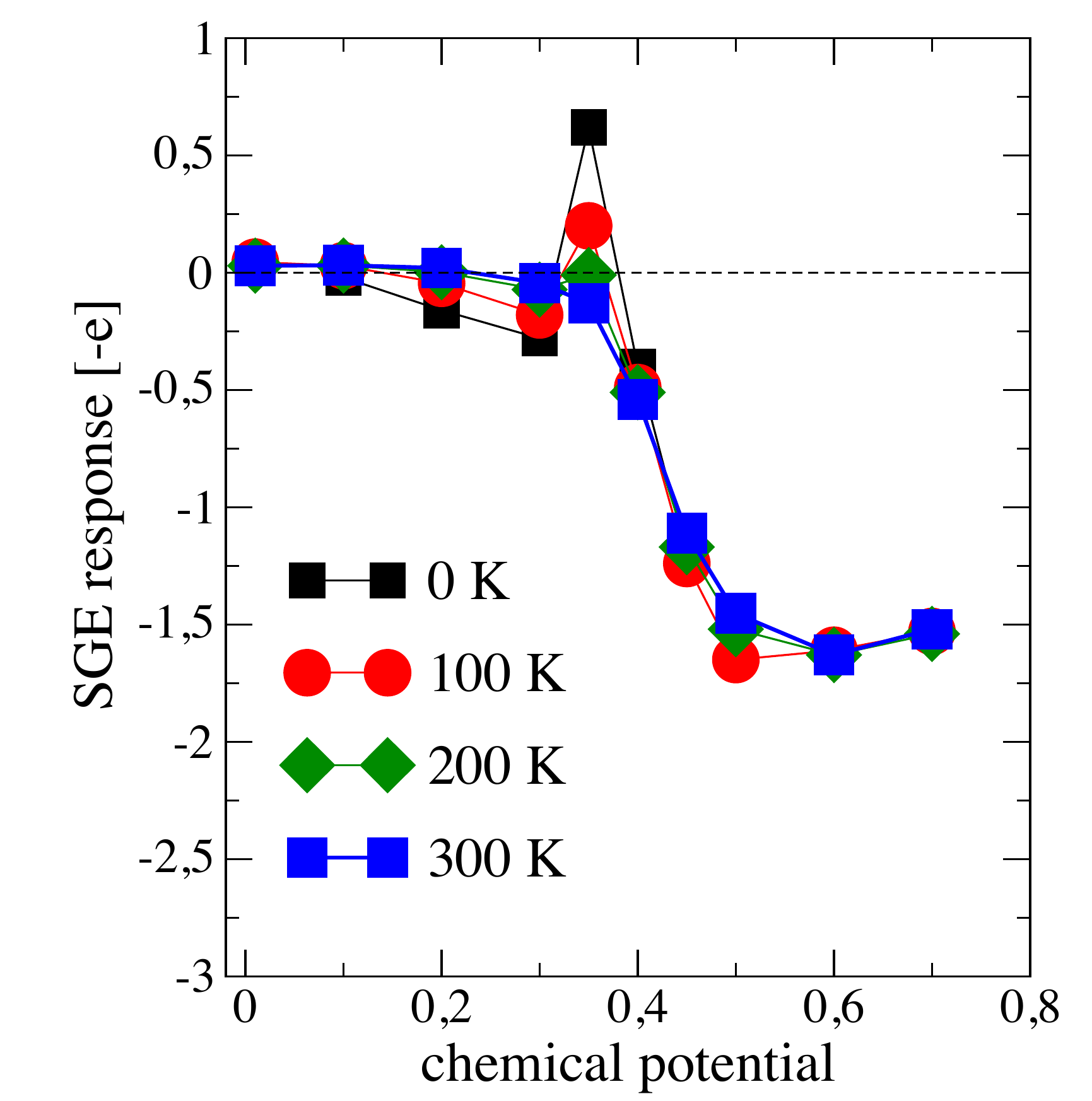}
\end{center}
\caption{Regular part of the SGC response as a function of chemical potential.
Disorder potential is $V_{0}=0.1$ eV and calculations are performed
on 24 \texttimes{} 24 lattices.
%Inset: Fit of the low energy optical
%conductivity to the Drude model for two chemical potentials $\mu=0.2$
%\textgreek{m} eV (circles) and $\mu=0.6$ eV yielding transport scattering
%times $\tau=0.03$ ps and $\tau=0.01$ ps, respectively.
\label{fig:sigmavdis}}
\end{figure}

Fig.\,\ref{fig:sigmavdis} shows the SGC of the disordered system
for four different temperatures, obtained by averaging over 50 disorder
configurations and over $100$ phase pairs $(\Phi_{x},\Phi_{y})$
for each disorder realization. To estimate the effective strength
of the disorder, we have evaluated the frequency-dependent electrical
longitudinal conductivity, whose Lorentzian lineshape allows to extract
the eleastic scattering time $\tau$, used in the analytical theory
of section \ref{sec:disorderedlimit}. For two chemical potentials
$\mu=0.2$ eV and $\mu=0.6$ eV, below and above the Lifshitz point,
the estimated scattering time is of the order of $10^{-2}$ ps, which
corresponds to a level broadening of the order of $10^{-5}$ eV. In
the presence of the SOC a crucial parameter is the ratio between the
spin-orbit split gap and the disorder-induced broadening. Keeping
in mind the typical size of the spin-orbit split gap shown in Fig.
\ref{fig:figapp}, one may conclude that the condition of weak scattering
limit is satisfied. At zero temperature, the black line in Fig. \ref{fig:sigmavdis}
shows that the SGC changes sign twice. One sees that the two sign
changes occur in a very restricted range of chemical potentials, when
first the pair of bands (3,4) starts to be occupied and then also
the pair of bands (5,6) becomes occupied as well. One then is tempted
to associate the positive sign with the initial filling of bands (3,4)
and the negative sign with the filling of bands (5,6) in agreement
with the analytical results of Eqs.(\ref{eq:Edelsteinm}) and (\ref{eq:Edelsteinp}).
The effect of the temperature reduces the value of the SGC. This happens
when the energy scale associated with the temperature becomes larger
than disorder broadening, which is the situation already at $100$
K. At finite temperature the SGC is likely to be an effective average
over its value at different chemical potentials, and hence over the
values associated to the different pairs of bands. As a result, at
the highest temperature $300$ K, there is only one sign change before
the Lifshitz point. Previously\cite{Seibold2017} it has been noticed
that the behavior at $T=300$ K is compatible with the experimental
behavior of Ref.\cite{Song2017}, whereas the sign change upon voltage
reversal of the experiment of Ref.\cite{Lesne2016}, performed at
$T=7$ K, can be interpreted as the second sign change of our $T=0$
K curve.

\section{CONCLUSIONS}

In this paper we have presented a detailed theoretical investigation
of the spin galvanic effect in a multi-band model describing the electron
states at a LAO/STO metallic interface. Starting from a tight-binding
description, we have derived a low-energy continuum model, which well
describes the original model close to the $\Gamma$ point.
The resulting effective Rashba-like models correspond to
a linear-in-momentum SOC  for the lowest and highest pair of bands while
it is cubic for the middle pair of bands. 
For these
effective models we have performed analytical calculations both in
the absence and in the presence of disorder.
In particular, we
have used the standard diagrammatic approach of impurity technique
valid in the metallic regime. We have also performed exact numerical
calculations, which are in agreement with the analytical ones close
to the $\Gamma$ point. The main results can be summarized as follows.
1) In the absence of disorder, the SGC as a function of frequency
of the driving electric field has a singular delta-like behavior reminiscent
of the Drude peak in the standard optical electrical conductivity.
The spectral strength associated to the delta function gets contributions
from all the interband transitions and, in general, cannot simply
attributed to a single pair of spin-orbit split bands. 2) The frequency-dependent
SGC has also a regular contribution, which in the absence of disorder
vanishes exactly at zero frequency. This regular part has a number
of spectral features, whose associated frequencies correspond to the
possible interband transitions. 3) A generic level-broadening mechanism
leads to a finite regular part at low frequency, whose behavior is
then dominated by the smallest energy interband transition. The latter
then can be directly linked to a specific pair of spin-orbit split
bands. A numerical calculation inevitably requires a finite level
broadening and we have shown the effect of varying the size of the
broadening. 4) The presence of disorder guarantees a stationary solution
and introduces an intrinsic level broadening, whose effective strength
we have estimated by looking at the Lorentzian lineshape of the electrical
conductivity as function of frequency.  Note that in contrast to the
SGC, the spin Hall effect for a Rashba model with linear coupling (as for
the lowest xy-type bands) would vanish \cite{Raimondi2005} under stationary conditions
and can only be sustained under special conditions, as e.g. a periodic
modulation of the chemical potential \cite{Seibold2015}.
5) The behavior of the SGC
as a function of the chemical potential shows a non monotonous behavior
at zero temperature, which evolves to a monotonous one when the temperature
becomes larger than the level broadening. 6) Our theoretical results
are compatible with recent experiments and call for a systematic study
of the voltage dependence as a function of the temperature.

\acknowledgments % equivalent to \section*{ACKNOWLEDGMENTS}       

G. S. acknowledges support from the Deutsche Forschungsgemeinschaft
under SE806/19-1.
S. C.  acknowledge financial support from the University of
Rome Sapienza Research Project No. RM116154AA0AB1F5.

\appendix
%dummy comment inserted by tex2lyx to ensure that this paragraph is not empty%dummy comment inserted by tex2lyx to ensure that this paragraph is not empty

\section{THE BETHE--SALPETER EQUATION FOR THE CHARGE CURRENT VERTEX}

\label{sec:BSVertex}

In this appendix we provide a few details on the solution of the Bethe--Salpeter
equation for the vertex. We follow closely the discussion developed
for the case of the Rashba 2DEG model\cite{Schwab2002}.

\subsection{The case of the $E^{xy}$ bands}

We begin with the case of the lowest pair of bands due to the $d_{xy}$
orbitals. This case is practically equivalent to the standard Rashba
2DEG model. Since vertex corrections do not modify the momentum dependence
of the vertex, it us useful to write the full vertex as

\begin{equation}
\tilde{J}_{x}=(-e)\frac{k_{x}}{m}\tau^{0}+\varGamma_{x},\label{eq:app1}
\end{equation}
where all the momentum dependence is limited to the \emph{bare} vertex
$(-e)k_{x}/m$. The spin-dependent part of the vertex\emph{ $\Gamma_{x}$
}satisfies then a new Bethe--Salpeter equation

\begin{equation}
\Gamma_{x}=\gamma_{x}+n_{i}u^{2}\sum_{\mathbf{k}}G^{R}\Gamma_{x}G^{A},\label{eq:app2}
\end{equation}
where the effective bare vertex is defined by

\begin{equation}
\gamma_{x}=(-e)\alpha_{xy}\tau^{y}+n_{i}u^{2}\sum_{\mathbf{k}}G^{R}(-e)\frac{k_{x}}{m}G^{A}.\label{eq:app3}
\end{equation}
In the above we have omitted for the sake of simplicity the explicit
frequency and momentum dependence of the Green functions. To evaluate
the integral over the momentum, one must use the Pauli matrix expansion
of the Green function shown in Eq.\,(\ref{eq:PauliGreenFunction}).
Because of the factor $k_{x}$ in the integral, only the combination
$G_{0}^{R}G_{2}^{A}$ and its complex conjugate appear. As a result
the integral in the right hand side of Eq.\,(\ref{eq:app3}) is proportional
to $\tau^{y}$ and exactly cancels the first term so that the vertex
$\gamma_{x}$ vanishes (see Ref. \cite{Raimondi2005} for details)
and the full vertex reduces to the standard current vertex as shown
in Eq.\,(\ref{eq:dressedvertexxy}).

\subsection{The case of the $E^{-}$ bands}

We follow the same strategy as in the previous case. The Green function
has now the form (we omit the frequency and momentum dependence for
brevity)

\begin{equation}
\hat{G}=\frac{G_{+}+G_{-}}{2}-\frac{\zeta}{|\zeta|}(\tau^{x}\hat{k}_{y}-\tau^{y}\hat{k}_{x})\frac{G_{+}-G_{-}}{2},\label{eq:app4}
\end{equation}
where $\zeta$ was introduced in Eq.\,(\ref{eq:Eigenem}). In this
case the effective bare vertex reads

\begin{equation}
\gamma_{x}=(-e)n_{i}u^{2}\sum_{\mathbf{k}}G^{R}\left(\frac{k_{x}}{m}\tau^{0}-2\beta k_{x}k_{y}\tau^{x}+\beta(3k_{x}^{2}-k_{y}^{2})\tau^{y}\right)G^{A}=(-e)\frac{1}{4}\beta p_{F}^{2}\tau^{y},\label{eq:app5}
\end{equation}
which must be inserted in Eq.\,(\ref{eq:app2}) with the form of
the Green functions given by Eq.\,(\ref{eq:app4}). In the above
$p_{F}$ is the Fermi momentum in the absence of SOC. Given the form
(\ref{eq:app5}), we look for a solution of the form $\Gamma_{x}=\Gamma_{x}^{y}\tau^{y}$.
With this ansatz, one easily sees that the integral over the momentum
in Eq.\,(\ref{eq:app2}) yields a term proportional to $\tau^{y}$.
As a result one has the closed equation

\begin{equation}
\Gamma_{x}^{y}=\gamma_{x}^{y}+I\Gamma_{x}^{y}=\frac{\gamma_{x}^{y}}{1-I},\label{eq:app6}
\end{equation}
where

\begin{equation}
I=n_{i}u^{2}\sum_{\mathbf{k}}\frac{1}{2}\mathrm{Tr}\left[\tau^{y}G^{R}\tau^{y}G^{A}\right]=1-\frac{1}{2}\left\langle \frac{4\beta^{2}p_{F}^{6}\zeta^{2}\tau^{2}}{1+4\beta^{2}p_{F}^{6}\zeta^{2}\tau^{2}}\right\rangle ,\label{eq:app7}
\end{equation}
where $\left\langle \ldots\right\rangle $ stands for the angle average
over the direction of momentum. In the weak disorder limit, $\tau\rightarrow\infty$,
$I=1/2$. As a result $\Gamma_{x}^{y}=(-e)\beta p_{F}^{2}/2$.

\subsection{The case of the $E^{+}$ bands}

In this case the Green function reads

\begin{equation}
\hat{G}=\frac{G_{+}+G_{-}}{2}+(\tau^{x}\hat{k}_{y}+\tau^{y}\hat{k}_{x})\frac{G_{+}-G_{-}}{2}.\label{eq:app8}
\end{equation}
The evaluation of the effective bare vertex is similar to the case
of the $E^{xy}$ bands with the replacement $\alpha_{xy}\rightarrow\alpha_{+}$.
As a result $\gamma_{x}=0$ and the dressed vertex coincides with
the momentum depedent part of the bare vertex.

\section{THE BETHE--SALPETER EQUATION AT FINITE FREQUENCY}

\label{sec:FFBS}

For the Rashba 2DEG model (\ref{eq:Rashba2DEG}), the Bethe--Salpeter
equation at finite frequency reads

\begin{eqnarray}
\Gamma_{x} & = & \gamma_{x}+n_{i}u^{2}\sum_{{\bf k}}\frac{1}{2}{\rm Tr}\Big\{\tau^{y}G_{{\bf k}}^{R}(\omega/2)\tau^{y}\Gamma_{x}G_{{\bf k}}^{A}(-\omega/2)\Big\}\label{gamma_eq}\\
\gamma_{x} & = & e\alpha\tau^{y}+n_{i}u^{2}\sum_{{\bf k}}\frac{1}{2}{\rm Tr}\Big\{\tau^{y}G_{{\bf k}}^{R}(\omega/2)(-e)\frac{k_{x}}{m}G_{{\bf k}}^{A}(-\omega/2)\Big\}.\label{eff_ver}
\end{eqnarray}
which has the solution

\[
\Gamma_{x}=-e\frac{\gamma_{x}}{\tau/\tau_{s}-{\rm i}\omega\tau},\ \ \gamma_{x}=(-e)\alpha\frac{{\rm i}\omega}{-{\rm i}\omega+1/\tau}\tau^{y}.
\]
The dressed vertex reads then

\begin{equation}
\tilde{J}_{x}=(-e)\frac{k_{x}}{m}\tau^{0}+-e\frac{1}{\tau/\tau_{s}-{\rm i}\omega\tau}\alpha\frac{{\rm i}\omega}{-{\rm i}\omega+1/\tau}\tau^{y}.\label{eq:DressedFFVertex}
\end{equation}
When the full dressed vertex (\ref{eq:DressedFFVertex}) is used in
the Kubo formula (\ref{eq:KuboGF}) one obtains Eq.\,(\ref{eq:LorentzianDrude}).

\end{document}